\documentclass[lettersize,journal]{IEEEtran}
\usepackage{amsmath,amsfonts}
\usepackage{algorithmic}
\usepackage{algorithm}
\usepackage[algo2e]{algorithm2e} 
\usepackage{graphicx}
\newtheorem{dfn}{Definition}
\newtheorem{lm}{Lemma}
\newcommand\independent{\protect\mathpalette{\protect\independenT}{\perp}}
\def\independenT#1#2{\mathrel{\rlap{$#1#2$}\mkern2mu{#1#2}}}

\begin{document}
\title{\textit{CURATE}: Scaling-up Differentially Private Causal Graph Discovery
\\ \thanks{This work was supported by NSF grants CAREER 1651492, CCF-2100013, CNS-2209951, CNS-1822071, CNS-2317192, and by the U.S. Department of Energy, Office of Science, Office of Advanced Scientific Computing under Award Number DE-SC-ERKJ422, and NIH Award R01-CA261457-01A1. Parts of this work were presented in IEEE International Workshop on Machine Learning for Signal Processing 2024 (IEEE MLSP 2024).}}
 \author{%
   \IEEEauthorblockN{Payel Bhattacharjee~~~ Ravi Tandon}
  \\
  {Department of Electrical and Computer Engineering\\
                     University of Arizona, Tucson, AZ, USA.\\
                     E-mail: {\{\textit{payelb, tandonr}\}}@arizona.edu
                   }
}
%


\maketitle
\begin{abstract}
    Causal Graph Discovery (CGD) is the process of estimating the underlying probabilistic graphical model that represents joint distribution of features of a dataset. CGD-algorithms are broadly classified into two categories: (i) Constraint-based algorithms (outcome depends on conditional independence (CI) tests), (ii) Score-based algorithms (outcome depends on optimized score-function). Since, sensitive features of observational data is prone to privacy-leakage, Differential Privacy (DP) has been adopted to ensure user privacy in CGD. Adding same amount of noise in this sequential-natured estimation process affects the predictive performance of the algorithms. As initial CI tests in constraint-based algorithms and later iterations of the optimization process of score-based algorithms are crucial, they need to be more accurate, less noisy. Based on this key observation, we present \textit{CURATE} (CaUsal gRaph AdapTivE privacy), a DP-CGD framework with adaptive privacy budgeting. In contrast to existing DP-CGD algorithms with uniform privacy budgeting across all iterations, \textit{CURATE} allows adaptive privacy budgeting by minimizing error probability (for constraint-based), maximizing iterations of the optimization problem (for score-based) while keeping the cumulative leakage bounded. To validate our framework, we present a comprehensive set of experiments on several datasets and show that \textit{CURATE} achieves higher utility compared to existing DP-CGD algorithms with less privacy-leakage.
\end{abstract}


\section{Introduction}

\label{intro}Causal graph discovery (CGD) enables the estimation of the partially connected directed acyclic graph (DAG) that represents the underlying joint probability distribution of the features of the observational dataset. CGD is an important part of causal inference \cite{spirtes_causation_1993} and is widely used in various disciplines, including biology \cite{sachs_causal_2005}, genetics \cite{zhang_integrated_2013}, drug discovery, ecology, criminal justice reform, curriculum design, finance and banking sectors.
\newline
\textit{\textbf{Overview of  Causal Graph Discovery (CGD): }} The estimation process of the causal graph from observational data relies on the execution of the causal graph discovery algorithms.
The CGD-algorithms are broadly classified into two categories: \textit{constraint-based algorithms} and \textit{score-based algorithms}. Constraint-based algorithms including PC \cite{spirtes_causation_1993}, FCI (Fast Causal Inference) \cite{spirtes_anytime_2001} and their variants \cite{nogueira_methods_2022} estimate the causal graph in two phases: first, the \textit{skeleton phase} in which the algorithm starts with a fully connected graph, and based on the statistical conditional independence (CI) test results, it updates the graph and returns a partially connected undirected graph. To determine conditional independence, a variety of test statistics, such as \textit{G-test }\cite{mcdonald_handbook_nodate}, $\chi^2$-\textit{test} \cite{mchugh_chi-square_2013}, correlation coefficients including \textit{Kendall's Tau} \cite{kendall_new_1938}, \textit{Spearman's Rho }\cite{spearman_proof_1987} can be used. In the second phase, \textit{orientation phase,} the algorithm orients the undirected edges based on the CI test results obtained in the skeleton phase and returns the estimated causal graph. The constraint-based algorithms theoretically guarantee to converge to the complete partial directed acyclic graph (CPDAG) under certain conditions including the correctness of the CI tests, causal sufficiency, Markov assumptions, etc. On the other hand, the score-based algorithms estimate the causal graphs from observational datasets by optimizing a score function. The algorithm essentially assigns relevance scores such as Bayesian Dirichlet equivalent uniform (BDe(u)\cite{bdeu}), Bayesian Gaussian equivalent (BGe\cite{bge}), Bayesian Information Criterion (BIC \cite{BIC}), and Minimum Description Length (MDL \cite{mdl}) to all the potential candidate graphs derived from the dataset and estimates the best graph out of them. This method enables the score-based algorithms to eliminate the necessity of a large amount of CI tests.  The recent work, NOTEARS \cite{zheng_dags_2018} proposes the idea of converting the traditional combinatorial problem to a continuous optimization problem in order to estimate the DAG. These algorithms, however, are computationally more expensive since they must enumerate and score each and every conceivable graph among the variables provided.
\newline
\textit{\textbf{Privacy Threats and Differentially Private CGD: }} CGD algorithms often deal with real-world datasets which may contain sensitive and private information about the participants including social and demographical information, credit history, medical conditions and many more. Releasing the causal graph itself or the intermediate statistical conditional independence (CI) test results often leads to the problem of privacy leakage. Recent work \cite{murakonda2021quantifying} demonstrates the \textit{membership inference threats} through probabilistic graphical models.
Several recent works adopt the notion of Differential Privacy (DP) \cite{dwork_our_2006} in the context of CGD to ensure a certain level of user privacy. 
\begin{figure*}
    \centering
    \includegraphics[scale = 0.27]{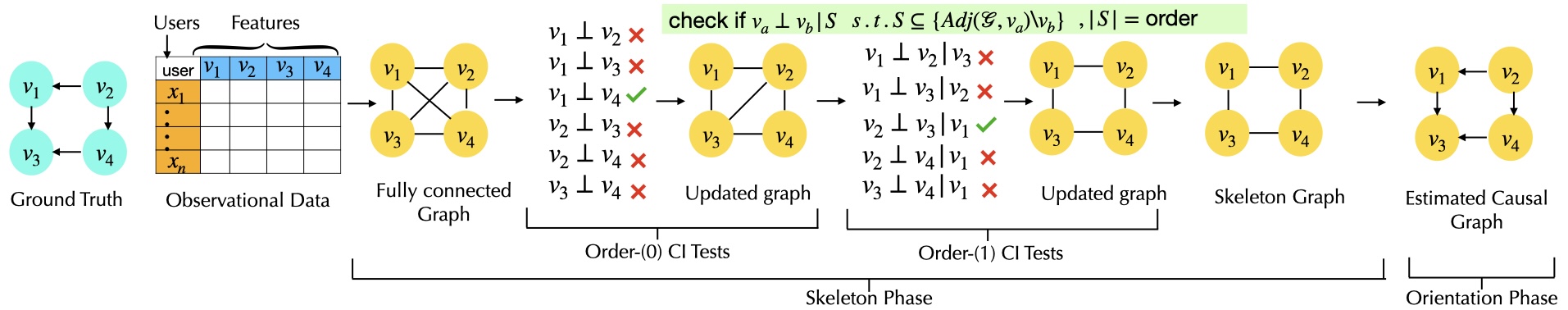}
    \caption{The generic workflow of constraint-based CGD algorithms with two phases: Skeleton Phase and Orientation Phase. The skeleton phase starts with a fully connected graph with $d$ nodes, where $d$ is the number of features/variables. $k_i$ is the maximum number of CI tests in order $i$. The sequence and number of tests in any order $i$ are dependent on the outcomes of order $(i-1)$ tests, and the skeleton phase is prone to privacy leakage.}
    \label{fig1_workflow}
\end{figure*}
For instance, the existing constraint-based differentially private CGD (DP-CGD) algorithms incorporates several differential privacy techniques to perturb the CI test statistic such as \textit{Laplace Mechanism} (PrivPC) \cite{wang_towards_2020}, \textit{Exponential Mechanism} (EM-PC) \cite{xu_differential_2017}, \textit{Sparse Vector Technique} (SVT-PC) \cite{wang_towards_2020}. For the class of score-based algorithms, NOLEAKS \cite{ma_noleaks_2022} adopts \textit{Gaussian Mechanism} to perturb the gradient of the optimization problem. However, it is observed that the existing algorithms rely on the method of adding the \textit{same amount of noise} to each iteration of the estimation process. As shown in Figure \ref{fig1_workflow} and discussed in Section \ref{sec:CURATE}, the CI tests in constraint-based CGD can be highly interdependent. If an edge between two variables is deleted by a CI test, then the conditional interdependence between them (conditioned on any other subset of features) is never checked in later iterations. Furthermore, this issue also impacts the scalability of private CGD; the total privacy leakage blows up for datasets with a large number of features ($d>>1$). 
Meanwhile, the differentially private score-based algorithms such as NOLEAKS \cite{ma_noleaks_2022} optimize objective function to obtain the adjacency matrix of estimated DAG. This optimization technique utilizes noisy gradients of the objective function, and adding the same amount of noise may leads to higher convergence time as the optimal point may be missed by the algorithm during the noise addition. In order to prevent the algorithm from missing the optima and make the converge faster, the later iterations of the optimization process should ideally be less noisy.
\newline
\textit{\textbf{Overview of the proposed framework CURATE: }}The aforementioned observations bring forth the important point of adaptive privacy budgeting for both constraint-based and score-based differentially private CGD algorithms. For constraint-based algorithms, the initial CI tests and for score-based algorithms, the later iterations in the optimization are more critical. This motivates the idea of \textit{adaptive privacy budgeting} given a total privacy budget which can reduce the risk of error propagation to subsequent iterations, and also improve the scalability of constraint-based algorithms. On the other hand, score-based algorithms ideally require less noise and more accuracy for the later iterations. Intuitively, higher privacy budget allocation to later iterations of the optimization process helps to reduce the risk of missing the optima of the objective function.
In this paper, we present an adaptive privacy budgeting framework \textit{CURATE} (CaUsal gRaph AdapTivE privacy) for both constraint-based and score-based CGD algorithms in a differentially private environment. The main contributions of this paper are summarized as follows:  
\begin{itemize}
    \item Our proposed framework \textit{CURATE} scales up the utility of the CGD process by adaptive privacy budget allocation. For the scope of constraint-based DP-CGD algorithms, constraint-based \textit{CURATE} algorithm optimizes privacy budgets for each order of CI test (CI tests of same order have same privacy budget) in a principled manner with the goal of minimizing the surrogate for the total probability of error. By allocating adaptive (and often comparatively higher) privacy budgets to the initial CI tests, \textit{CURATE}  ensures overall better predictive performance with less amount of total leakage compared to the existing constraint-based DP-CGD algorithms. 
    \item We present score-based \textit{CURATE} algorithm which allows adaptive budgeting that maximizes the iterations given a fixed privacy budget ($\epsilon_{\text{Total}}$) for the scope of score-based algorithms. The score-based \textit{CURATE} algorithm uses functional causal model based optimization approach that allocates a higher privacy budget to the later iterations. As the privacy budget gets incremented as a function of iterations, score-based \textit{CURATE} achieves better utility compared to the existing work(s).
    \item In this paper, we present extensive experimental results on 6 public CGD datasets. We compare the predictive performance of our proposed framework \textit{CURATE} with existing DP-CGD algorithms. Experimental results show that \textit{CURATE} ensures better predictive performance with leakage smaller by orders of magnitude. The average required CI tests in constraint-based \textit{CURATE} is also significantly less than the existing constraint-based DP-CGD algorithms.
\end{itemize}
\section{Preliminaries on CGD and DP}\label{prelim}
In this Section, we review the notion of causal graph discovery and provide a brief overview of both constraint-based algorithms (canonical PC algorithm) and FCM-based algorithms (NOTEARS, NOLEAKS algorithms) along with the description of differential privacy \cite{dwork_our_2006,dwork_calibrating_2006}.
\begin{dfn}[Probabilistic Graphical Model] Given a joint probability distribution $\mathbb{P}(F_1,\ldots,F_d)$ of $d$ random variables, the graphical model $\mathcal{G}^*$  with $V$ vertices ($v_1,\ldots,v_d$) and $E\subseteq V\times V$ edges is known as \textit{Probabilistic Graphical Model (PGM)} if the joint distribution decomposes as:
\begin{align*}
    \mathbb{P}(F_1,\ldots,F_d)=\prod_{F_a\in\{F_1,\ldots,F_d\}} \mathbb{P}(F_a|Pa(F_a)),
\end{align*} 
where, $Pa(F_a)$ represents the direct parents of the node $F_a$. It relies on the assumption of  probabilistic independence ($F_a\independent_P F_b|S$) $\implies$ graphical independence ($v_a\independent_G v_b|S$) \cite{zanga2022survey}.   
\end{dfn}
\begin{dfn}[Causal Graph Discovery] Given dataset $\mathcal{D}$ with the collection of $n$ i.i.d. samples $(\mathbf{x_1},\ldots, \mathbf{x_n})$ drawn from a joint probability distribution $\mathbb{P}(F_1,\ldots,F_d)$ where $\mathbf{x}_{i}$ is a $d$-dimensional vector representing the $d$ features/variables of the $i^{th}$ sample (user); the method of estimating the \textit{PGM} $(\mathcal{G}^*)$ from $\mathcal{D}$ is known as \textit{Causal Graph Discovery (CGD)}\cite{wang_towards_2020}.
\end{dfn}
\begin{dfn}[($\epsilon,\delta$)-Differential Privacy] 
\cite{dwork_our_2006,dwork_calibrating_2006,dwork_algorithmic_2013} For all pair of neighboring datasets $\mathcal{D}$ and $\mathcal{D}'$ that differ by a single element, i.e., $||\mathcal{D}-\mathcal{D}'||_1\le 1$, a randomized algorithm $\mathcal{M}$ with an input domain of D and output range $\mathcal{R}$ is considered to be $(\epsilon,\delta)$-\textit{differentially private}, if $\forall \mathcal{S} \subseteq \mathcal{R}$:
\begin{align*}
    \mathbb{P}[\mathcal{M}(\mathcal{D})\in \mathcal{S}]\le e^{\epsilon}\mathbb{P}[\mathcal{M}(\mathcal{D}')\in \mathcal{S}]+\delta.
\end{align*}
\end{dfn}
Differentially private CGD algorithms have adopted \textit{Exponential Mechanism} \cite{xu2017differential}, \textit{Laplace Mechanism, Sparse Vector Technique} \cite{wang_towards_2020}, \textit{Gaussian Mechanism} \cite{ma_noleaks_2022} to ensure DP. 
\begin{dfn}[$l_k$- sensitivity]
For two neighboring datasets $\mathcal{D}$ and $\mathcal{D}'$, the $l_k$-sensitivity of a function $f(\cdot)$ is defined as:
\begin{align*}
    \Delta_k(f) = \max_{\mathcal{D},\mathcal{D}'\in \mathcal{R},|\mathcal{D},\mathcal{D}'|\le 1} ||f(\mathcal{D})-f(\mathcal{D}')||_k.
\end{align*}
\end{dfn}
For instance, Laplace mechanism perturbs the CI test statistic $f(\cdot)$ with Laplace noise proportional to the $l_1$-sensitivity of the function $f(\cdot)$, whereas Gaussian mechanism adds noise proportional to the $l_2$-sensitivity to ensure DP-guarantee.
Ideally, the Classical Gaussian Mechanism uses $\epsilon\le1$ for ($\epsilon,\delta$)-DP guarantees, however, this condition may not be sufficient in all scenarios of CGD \cite{ma_noleaks_2022}. Therefore, the DP score-based algorithm \cite{ma_noleaks_2022} uses Analytical Gaussian Mechanism \cite{balle_privacy_2018}.
\begin{dfn}[Analytic Gaussian Mechanism \cite{balle_privacy_2018}]
For a function $f:\mathbb{X}\leftarrow \mathbb{R}^d$ with $l_2$-sensitivity $\Delta_2$ and privacy parameters $\epsilon\ge 0$ and $\delta\in [0,1]$, the Gaussian output perturbation mechanism $\mathcal{A}(x)=f(x)+Z$ with $Z~ \mathcal{N}(0,\sigma^2 I)$ is $(\epsilon,\delta)$-DP if and only if:
   \begin{equation}\label{sigma}
     \Phi\left(\frac{\Delta_2}{2\sigma}-\frac{\epsilon \sigma}{\Delta_2}\right)-e^{\epsilon}\Phi\left(\frac{-\Delta_2}{2\sigma}-\frac{\epsilon \sigma}{\Delta_2}\right)\le \delta,  
   \end{equation}
where $\Phi$ is the CDF of the Gaussian Distribution.
\end{dfn}
\textit{\textbf{Overview of Constraint-based Algorithms: }}
Canonical constraint-based CGD algorithms (such as the PC algorithm  \cite{spirtes_causation_1993}) work in two phases: a \textit{skeleton phase} followed by an \textit{orientation phase}. In the \textit{skeleton phase}, the algorithm starts with a fully connected graph ($\mathcal{G}$) and prunes it by conducting a sequence of conditional independence (CI) tests.  
The CI tests in PC are order dependent, and the order of a test represents the cardinality of the conditioning set $S$ of features. In order-$(i)$ tests, all the connected node pairs ($v_a,v_b$) in $\mathcal{G}$ are tested for statistical independence conditioned on the set $S$. The conditioning set $S$ is chosen such that  $S\subseteq \{Adj(\mathcal{G},v_a)\text{\textbackslash{}} v_b\}$, where $Adj(\mathcal{G},v)$ represents the adjacent vertices of the node $v$ in the graph $\mathcal{G}$. Edge between the node pairs ($v_a,v_b$) gets deleted if they pass order-$(i)$ CI test and never get tested again for statistical independence conditioned on set $S$ with $|S|>i$. The remaining edges in $\mathcal{G}$ then get tested for independence in order-$(i+1)$ CI tests conditioned on a set $S$ with $|S|=(i+1)$. This process of CI testing continues until all connected node pairs in $\mathcal{G}$ are tested conditioned on  set $S$ of size ($d-2$). At the end of this phase, PC returns the skeleton graph. In the \textit{orientation phase}, the algorithm orients the edges based on the separation set $S$ of one independent node pair ($v_a,v_b$) without introducing cyclicity in $\mathcal{G}$ \cite{spirtes_causation_1993,xu_differential_2017} as shown in Figure \ref{fig1_workflow}. The privacy leakage in this two-step process is only caused in the \textit{skeleton phase}, as this is when the algorithm directly interacts with the dataset $\mathcal{D}$.  
Thus, privacy leakage in this two-step process is only caused in the skeleton phase, as this is when the algorithm directly interacts with the dataset $\mathcal{D}$.  Therefore, the existing literature has focused on effectively privatizing CI tests subject to the notion of differential privacy \cite{dwork_our_2006,dwork_calibrating_2006} which ensures the presence/absence of a user will not \textit{significantly} change the estimated causal graph.
\newline
\textit{\textbf{Overview of Score-based Algorithms: }} Score-based algorithms estimate the DAG that optimizes a predefined score function. Due to the combinatorial acyclicity constraints, learning DAGs from data is NP-hard \cite{chickering1996learning}. To address this issue, the score-based CGD algorithm NOTEARS \cite{zheng_dags_2018} proposes a continuous optimization problem with an acyclicity constraint which estimates the DAG from observational data and eliminates the necessity of the search over the combinatorial space of DAGs.  From a group of DAGs, the one DAG is selected which optimizes a pre-defined score function $\mathtt{score}(\cdot)$ while satisfying the acyclicity constraints. Given an observational dataset $\mathcal{D}$ with $n$ i.i.d. samples and $d$-features $\mathcal{F}=(F_1,F_2,\ldots,F_d)$ the algorithm estimates (mimics) the data generation process $f_i(\cdot)$ for every $i^{th}$ feature/variable by minimizing the loss function. Essentially, the adjacency matrix $W$ that represents the edges of the graph $\mathcal{G}$ is modeled with the help Functional Causal Model (FCM). FCM-based methods represent every variable $i^{th}$ variable $F_i$  of the dataset $\mathcal{D}$ as a function of its parents $\text{Pa}(F_i)$ and added noise $Z$ as:
\begin{align*}
    F_i=f_i(\text{Pa}(F_i))+Z.
\end{align*}
The key idea behind FCM-based CGD is to estimate the weight vector $w_i$ for each variable $F_i$ given its parents $\text{Pa}(F_i)$.  Therefore each variable $F_i$ can be represented as a weighted combination of its parents and noise $Z$ as: $F_i=w_i^T\mathcal{F}+Z$.
The optimization process of estimating the weight vector $w_i$ is based on the idea of minimizing the squared loss function $\ell(W,\mathcal{D})=\frac{1}{2n}||\mathcal{D}-\mathcal{D}W||_F^2$, where $W$ is the associated adjacency matrix of the dataset $\mathcal{D}$ and $n$ is the number of samples. In the optimization process, the algorithm also uses a penalty function $\lambda||W||_1$ that penalizes dense graphs. The detailed working mechanism of FCM-based CGD algorithms is described in Section \ref{sec:CURATE}.
\newline
\textit{\textbf{Sensitivity Analysis and Composition of DP: }}For the class of constraint-based algorithms, an edge between the nodes ($v_a,v_b$) from estimated graph $\mathcal{G}$ gets deleted conditioned on set $S$ if ($f_{v_a,v_b|S}(\mathcal{D})> T$), where $f_{v_a,v_b|S}(\cdot)$ is the test statistic, and $T$ is the test threshold. Thus the structure of the estimated causal graph depends on the nature of $f(\cdot)$ and the threshold ($T$). Also, in DP-CGD, the amount of added noise is proportional to the $l_k$-sensitivity ($\Delta_k$) of the test statistic $f_{v_a,v_b|S}(\cdot)$. Therefore, to maximize the predictive performance, test statistics with lower sensitivity with respect to sample size $n$ are preferred. Through analysis we observed the $l_1$-sensitivity of the \textit{Kendall's} $\tau$ test statistic can be bounded as $\Delta_1\le \frac{C}{\sqrt{n}}$ ($C$ is a constant obtained from the analysis presented in \textit{Appendix \ref{appB}}). However, any other CI test statistics mentioned in Section \ref{intro} can be used in the framework of constraint-based \textit{CURATE}. The class of score-based algorithms focuses on the optimization of a score function to estimate the causal graph. Often these algorithms rely on gradient-based methods, and the gradient of the objective function frequently gets clipped and perturbed to preserve privacy. As mentioned in \cite{ma_noleaks_2022}, the $l_2$ sensitivity of the clipped gradient can be bounded as: $\Delta_2\le \frac{ds}{n}$ where $s$ is the clipping threshold. The paper \cite{ma_noleaks_2022} further exploits the properties of the dataset and adjacency matrix and the $l_2$ of the gradient is further upper-bounded as: $\Delta_2\le\frac{\sqrt{d(d-1)s}}{n}$.
\textit{Composition} is a critical tool in DP-CGD as the differentially private CGD algorithms discover the causal graph in an iterative process. Constraint based CGD algorithms run a sequence of interdependent tests, and score-based algorithms optimize the pre-defined score function in an iterative manner. Therefore the total leakage can be calculated by \textit{Basic Composition} \cite{dwork_our_2006,dwork_calibrating_2006,dwork_differential_2009,dwork_boosting_2010}, \textit{Advanced Composition} \cite{dwork_boosting_2010,dwork_algorithmic_2013}, \textit{Optimal Composition} \cite{kairouz_composition_2017}, \textit{Adaptive Composition} \cite{rogers_privacy_2016}, \textit{Moments Accountant} \cite{abadi_deep_2016}.
\section{Adaptive Differential Privacy in Causal Graph Discovery} \label{sec:CURATE} In this Section, we present the main idea of this paper, adaptive privacy budgeting framework \textit{CURATE}. In Section \ref{sec3.1}, we demonstrate the adaptive privacy budgeting mechanism for constraint-based algorithms. We introduce and explain the basic optimization problem that enables the allocation of the adaptive privacy budget through all the iterations (orders) of the CI tests. 
In Section \ref{sec3.2} we present adaptive privacy budget allocation for score-based algorithms. We introduce adaptivity while ensuring differential privacy (DP) during the evaluation of the weighted adjacency matrix. This section provides the theoretical foundation behind the adaptive privacy budget allocation mechanism in the context of DP-CGD.
\begin{figure*}[t]
    \centering
    \includegraphics[scale=0.27]{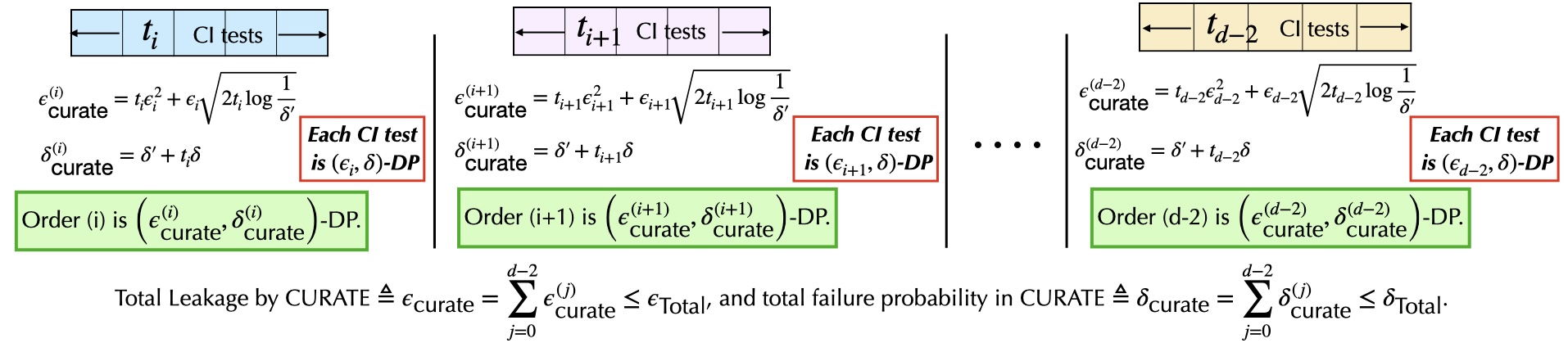}
    \caption{The composition mechanism in constraint-based \textit{CURATE} across all order of CI tests. For every order-(i), total privacy leakage is calculated with Advanced Composition sice the privacy budgets and failure probability for all order-$(i)$ tests are same. The total leakage across all orders is then calculated by constraint-based \textit{CURATE} with Basic Composition.}
    \label{curate_composition}
    
\end{figure*}
\subsection{Adaptive Privacy Budget Allocation with constraint-based \textit{CURATE} Algorithm: }\label{sec3.1} In this Section, we present the main proposed idea of this paper, \textit{CURATE}, that enables adaptive privacy budgeting while minimizing the error probability. As, the CI tests in constraint-based CGD algorithm are highly interdependent, predicting the total number of CI tests in CGD before the execution of the tests is difficult. The number of order-$(i)$ CI tests ($t_i$) enables the framework to have an approximation of per-order privacy budgets for later iterations ($\epsilon_{i},\ldots,\epsilon_{d-2}$) based on the total remaining privacy budget ($\epsilon_{\text{Total}}^{(i)}$). One naive data agnostic way to upper bound $t_i$ is: $t_i\le {d\choose 2}\cdot{d-2\choose i}$, where ${d\choose 2}$ represents the number of ways to select an edge from the edges of a fully connected graph (the way of selecting an edge between 2 connected nodes out of $d$ nodes), and ${d-2\choose i}$ refers to the selection of conditioning set ($S$) with cardinality $|S|=i$. However, this upper bound is too large and does not depend on the outcome of the previous iteration. A better approximation of $t_i$ is always possible given the outcome of the previous iteration. As, DP is immune to \textit{post-processing} \cite{dwork_calibrating_2006}, releasing the number edges ($e_{i+1}$) after executing order-$(i)$ differentially private CI tests will preserve differential privacy. For instance, the possible number of order-($i+1$) CI tests can always be upper-bounded as $t_{i+1}\le e_{i+1} \cdot{d-2 \choose i+1}$ where $e_{i+1}$ represents the remaining edges after order-$(i)$ tests. We have studied both of the methods and observed that $t_{i+1}\le e_{i+1}\cdot{d-2\choose i+1}$ is a better estimate of $t_{i+1}$ as $e_{i}\le {d\choose 2}, \forall i \in \{0,d-2\}$. Given the outcome of order-($i-1$) tests graph $\mathcal{G}$ with edges $e_i$ and a total (remaining) privacy budget of $\epsilon_{\text{Total}}^{(i)}$, we assign a privacy budgets $(\epsilon_i,\ldots,\epsilon_{d-2})$.
As every order-$(i)$ CI test in \textit{CURATE} is $(\epsilon_i,\delta)$-DP, with DP failure probabilities $\delta,\delta'>0$, the total leakage in order-$(i)$ is calculated with Advanced Composition \cite{dwork_algorithmic_2013} as: $\epsilon_\text{curate}^{(i)} = t_i\epsilon_i^2+\sqrt{2\log(\frac{1}{\delta'})t_i\epsilon_i^2}$, and the total failure probability in DP as: $\delta_{curate}^{(i)} =(\delta'+t_i\delta)$. However, as different orders have different privacy budgets, the total privacy leakage by \textit{CURATE} is calculated with Basic Composition \cite{dwork_algorithmic_2013} as: $\sum_{j=0}^{d-2}\epsilon_\text{curate}^{(j)}=\sum_{j=0}^{d-2}\left(t_j \epsilon_j^2 + \sqrt{2t_j\log(\frac{1}{\delta'}) \epsilon_j^2}\right)$, and the cumulative failure probability of \textit{CURATE} is $\sum_{j=0}^{d-2}\delta_{curate}^{(j)}$ (refer Figure \ref{curate_composition}). Therefore, given the outcome of order-$(i-1)$ tests, the total leakage in \textit{CURATE} must satisfy: 
$\sum_{j=i}^{d-2}\left(t_j \epsilon_j^2 + \sqrt{2t_j\log(\frac{1}{\delta'}) \epsilon_j^2}\right)\le \epsilon_{\text{Total}}^{(i)}$, where $t_j=e_j\cdot{d-2\choose j}$, and $\sum_{j=0}^{d-2}\delta_{curate}^{(j)}\le \delta_{\text{Total}}$. Moreover, we enforce $\epsilon_i\geq \epsilon_{i+1} \geq \ldots \geq \epsilon_{d-2}$, so that the initial CI tests get a higher privacy budget.  
\newline
\textbf{\textit{DP-CI Test in CURATE}}: The differentially private order-$(i)$ CI test with privacy budget $\epsilon_i$, for variables $(v_a,v_b)\in \mathcal{G}$ conditioned on a set of variables $S$ is defined as follows:
\begin{itemize}
    \item if $\hat{f} > T(1+\beta_2) \implies$ delete edge ($v_a,v_b$)
    \item else if $\hat{f} < T(1-\beta_1) \implies$ keep edge ($v_a,v_b$)
    \item else keep the edge with probability $\frac{1}{2}$,
\end{itemize}
where $\hat{f}:= f_{v_a,v_b|S}(\mathcal{D})+\text{Lap}(\frac{\Delta}{\epsilon_i})$, $\text{Lap}(\frac{\Delta}{\epsilon_i})$ is \textit{Laplace noise}, $\Delta$ denotes the $l_1$-sensitivity of the test statistic, $T$ denotes the threshold, and $(\beta_1, \beta_2)$ denote margins. In order to keep the utility high, one would ideally like to pick $(\epsilon_i, \epsilon_{i+1},\cdots, \epsilon_{d-2})$ that minimize the error probability $\mathbb{P}[E]=\mathbb{P}[\mathcal{G}\neq \mathcal{G}^*]$, where $\mathcal{G}^*$ is the true causal graph, and $\mathcal{G}$ is the estimated causal graph. Unfortunately, we do not have access to $\mathcal{G}^*$; in this paper, we instead propose to use a \textit{surrogate} for error by considering Type-I and Type-II errors relative to the unperturbed (non-private) statistic. Type-I error relative to the unperturbed CI test occurs when the private algorithm keeps the edge given that the unperturbed test statistic deletes the edge $(f_{v_a,v_b|S}(\mathcal{D})> T)$, and relative Type-II error occurs when the algorithm deletes an edge given that the unperturbed test statistic keeps that edge $(f_{v_a,v_b|S}(\mathcal{D})< T)$. The next \textit{Lemma} gives upper bounds on relative Type-I and Type-II error probabilities in \textit{CURATE}. 
\begin{lm}\label{cb_curate_lemma1}
For some $c_1, c_2 \in (0,1)$, and non-negative test threshold margins $(\beta_1,\beta_2)$, the relative Type-I ($\mathbb{P}[E_1^i]$) and Type-II ($\mathbb{P}[E_2^i]$) errors in order-$(i)$ CI tests in \textit{CURATE} with privacy budget $\epsilon_i$ and $l_1$-sensitivity $\Delta$ can be bounded as:
\begin{align*}
        \mathbb{P}[E_1^i]\le \underbrace{\frac{c_1}{2}+ \frac{1}{2}e^{(-\frac{T\beta_1\epsilon_i}{\Delta})}}_{q_i^{(1)}},
\quad
\mathbb{P}[E_2^i]\le \underbrace{\frac{c_2}{2}+ \frac{1}{2}e^{(-\frac{T\beta_2\epsilon_i}{\Delta})}}_{q_i^{(2)}}.
\end{align*}
\label{lemma1}
\end{lm}
\vspace{-2em}
The proof of \textit{Lemma 1} is presented in the \textit{Supplementary document}. The main objective of \textit{CURATE} is to allocate privacy budgets adaptively for order-$(i)$ CI tests by minimizing the total relative error. The leakage in DP-CGD depends on the number of CI tests and the number of CI tests depend upon the number of edges in the estimated graph $\mathcal{G}$. As, the number of edges in the true graph is not known, we use $\mathbb{P}[E_1^i]+\mathbb{P}[E_2^i]$ as a surrogate for the total error probability $\mathbb{P}[E]$. Given the outcome of order-($i-1$) tests, the algorithm can make Type-I error by preserving an edge which is not present in the true graph till order-($d-2$). If such an edge is present after order-($i-1$) tests, the probability of Type-1 error at the end of the order-($d-2$) can be represented as:  $\prod_{j=i}^{d-2}q_j^{(1)}$ since independent noise addition to each CI test enables the framework to bound the probability of error in each order independently and at the end of order-$(d-2)$ the total probability of error is the cumulative error made by the algorithm in every order-$(j)$. Similarly, probability of keeping an edge which is present in the ground truth after order-$(i-1)$ tests can be represented as $\prod_{j=i}^{d-2}(1-q_j^{(2)})$, therefore, the total Type-II error can be represented as: $\left(1-\left(\prod_{j=i}^{d-2}(1-q_j^{(2)})\right)\right)$. This leads to the construction of the main objective function  of this paper given the outcome of order-$(i-1)$ CI tests $\mathcal{G}$. The objective function that we propose to minimize is given as:
\begin{align}
\prod_{j=i}^{d-2}q_j^{(1)}+ \left(1-\left(\prod_{j=i}^{d-2}(1-q_j^{(2)})\right)\right).
    \label{edgeerror}
\end{align}
Since the number of edges in true graph are unknown, we propose to minimize \eqref{edgeerror} as a surrogate for the error probability.
\newline
\textit{\textbf{Optimization for Privacy Budget Allocation: }}By observing the differentially private outcome of order-($i-1$) CI tests (remaining edges $e_i$ in graph $\mathcal{G}$), \textit{CURATE} optimizes for $\Bar{\epsilon}=\{\epsilon_i,..,\epsilon_{d-2}\}$ (privacy budgets for subsequent order-$(i)$ tests and beyond) while minimizing the objective function as described in \eqref{edgeerror}. Formally, we define the optimization problem in \textit{CURATE}, denoted as $OPT(\epsilon_{\text{Total}}^{(i)},e_i,i)$:
\begin{align}
\normalsize
\label{prberreq}
\underbrace{\arg\min_{\Bar{\epsilon}} \prod_{j=i}^{d-2}q_j^{(1)}+ \left(1-\left(\prod_{j=i}^{d-2}(1-q_j^{(2)})\right)\right)}_{OPT(\epsilon_{\text{Total}}^{(i)},e_i,i)}\nonumber\\\text{ s.t. }
  \begin{cases}
        & \sum_{j=i}^{d-2}\underbrace{\left(t_j \epsilon_j^2 + \sqrt{2\log(\frac{1}{\delta'})t_j \epsilon_j^2}\right)}_{\text{total leakage in order-(j)}}\le \epsilon_{\text{Total}}^{(i)} \\
        &    \epsilon_j\ge \epsilon_{j+1}.
  \end{cases}
\end{align}
Given the outcome of order-$(i-1)$ tests, the above optimization function $OPT(\epsilon_{\text{Total}}^{(i)},e_i,i)$ takes the following inputs: (a)  remaining total budget ($\epsilon_{\text{Total}}^{(i)}$), (b) remaining edges ($e_{i}$) in the output graph $\mathcal{G}$ after all order-$(i-1)$ tests, (c) the index of order, i.e., $i$. The function then optimizes and outputs the privacy budgets ($\epsilon_{i},\ldots,\epsilon_{d-2}$) for remaining order tests, while satisfying the two constraints mentioned in \eqref{prberreq}. As the optimization problem in \eqref{prberreq} is difficult to solve in a closed form, in our experiments we have used Sequential Least Squares Programming (SLSQP) for optimizing the objective function.
\newline
\textit{\textbf{Constraint-based CURATE Algorithm}: } Now we present the constraint-based algorithm \textit{CURATE} that enables adaptive privacy budget allocation for each order-$i$ conditional independence tests by solving the optimization problem in  \eqref{prberreq}.

In constraint-based \textit{CURATE}, we use the optimization function $OPT(\cdot)$ recursively to observe adaptively chosen per-iteration privacy budgets. Given the total privacy budget for order-$i$ tests ($\epsilon_{\text{Total}}^{(i)}$), $OPT(\cdot)$ calculates the remaining privacy budget for order-$(i+1)$ CI tests based on $t_i$ number of order-$i$ CI tests:
\begin{align*}
\underbrace{\epsilon_{\text{Total}}^{(i+1)}}_{\text{budget for order-(i+1)}}=\underbrace{\epsilon_{\text{Total}}^{(i)}}_{\text{budget for order-i }}-\underbrace{\left(t_i\epsilon_i^2+\epsilon_i\sqrt{2t_i\log(\frac{1}{\delta'})}\right)}_{\text{actual leakage in order-$i$}}.
\end{align*}
Initially, the remaining budget for order-0 CI tests is equal to the assigned total privacy budget, i.e., $\epsilon_{\text{Total}}^{(0)}=\epsilon_{\text{Total}}$ and the edges in the complete graph $\mathcal{G}_0$ can be expressed as $e_0={d\choose 2}$. In order-0, \textit{CURATE} solves for ($\epsilon_0,\ldots,\epsilon_{d-2}$) by using the function $OPT(\epsilon_{\text{Total}}^{(0)},e_0,0)$. After completion of all order-$0$ CI tests, the algorithm calculates the remaining budget for order-1 CI tests as $\epsilon_{\text{Total}}^{(1)}=\epsilon_{\text{Total}}^{(0)}-\left(t_0\epsilon_0^2+\epsilon_0\sqrt{2t_0\log(\frac{1}{\delta'})}\right)$ and by observing the remaining edges $e_1$, it solves for the next set of privacy budgets ($\epsilon_1,\ldots,\epsilon_{d-2}$). 
We then recursively apply this process for all $i\in \{0,1,\ldots, d-2\}$ corresponding to all order-$i$ tests. 
\begin{figure*}[t]
    \centering
    \includegraphics[scale = 0.41]{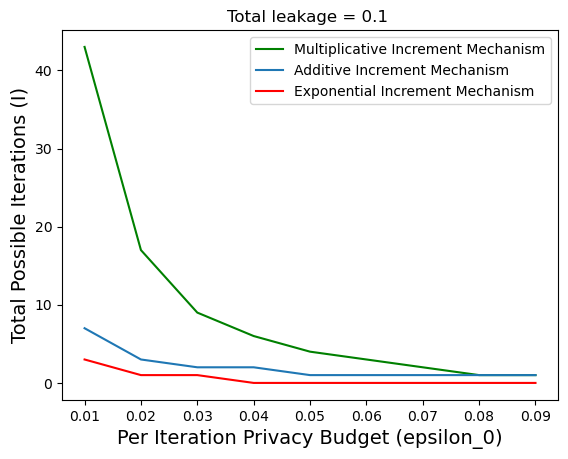}
    \includegraphics[scale = 0.41]{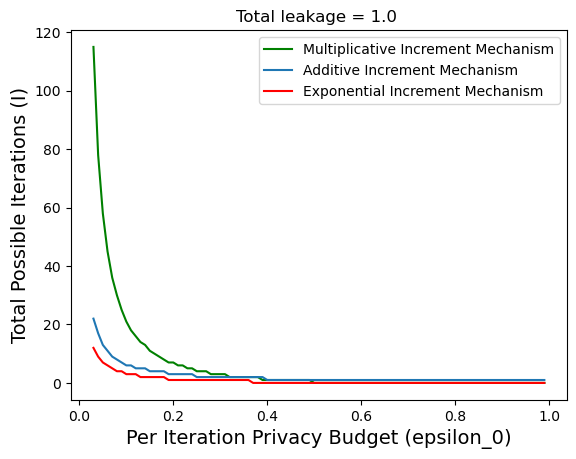}
    \includegraphics[scale = 0.41]{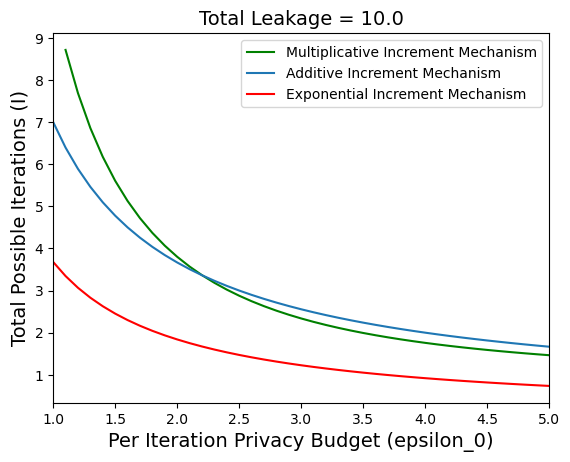}
    \caption{Possible number of iterations ($I$) given a total amount of privacy budget ($\epsilon_{\text{Total}}$) and initial privacy budget ($\epsilon_0$). For varied total privacy budget ($\epsilon_{\text{Total}}=0.1,\epsilon_{\text{Total}}=1.0$,$\epsilon_{\text{Total}}=10.0$) and different initial budget ($\epsilon_0<<1.0$ and $\epsilon_0>1.0$) we can observe that in the high privacy regime (i.e., $\epsilon_0<<1.0$) the multiplicative method executes more number of iterations.}
    \label{fig_adanl_methods}
    \vspace{-1em}
\end{figure*}
\textit{Sub-sampling} has also been adopted by several recent works on DP-CGD \cite{wang_towards_2020}, \cite{ma_noleaks_2022}. As \textit{sub-sampling} amplifies differential privacy \cite{balle_privacy_2018}, we can also readily incorporate sub-sampling parameters within the optimization framework of constraint-based and score-based \textit{CURATE}.
\begin{algorithm}[H]
\caption{\textit{CURATE Algorithm}}\label{online_CURATE_algo}
\KwData{Dataset $\mathcal{D}$, total privacy budget ($\epsilon_{\text{Total}}$), DP-failure probabilities $(\delta,\delta'>0)$, total failure probability ($\delta_{\text{Total}}$), test statistic $f(\cdot)$, threshold $T$, margins $(\beta_1,\beta_2)$, $l_1$-sensitivity $\Delta$, fully connected graph $\mathcal{G}$}
\KwResult{Partially connected graph $\mathcal{G}$}
Perform sub-sampling: $\mathcal{D}''\leftarrow^{\frac{m}{n}}\mathcal{D}, n = |\mathcal{D}|,m = |\mathcal{D}''|$\\
Initiation: $i=0$, $\epsilon_{\text{Total}}^{(0)}=\epsilon_{\text{Total}}$,  
$\delta\le 10^{-1.5m}$, $e_0={d\choose 2}$
\\
\For{$i = \{0,1,\ldots,d-2\}$}
{Initiate number of order-$i$ CI tests as: $t_i=0$\\
{
$(\epsilon_i,\ldots,\epsilon_{d-2})=OPT(\epsilon_{\text{Total}}^{(i)},e_i,i)$\\
$\forall $ connected node pairs $(v_a,v_b)$ in $\mathcal{G}$ that has not been tested on $S$ s.t. $S\subseteq \{Adj(\mathcal{G},v_a)\text{\textbackslash{}} v_b\},|S|=i$\\
Evaluate $\hat{f}:= f_{v_a,v_b|S}(\mathcal{D}'')+\text{Lap}(\frac{\Delta}{\epsilon_i})$\\
\begin{itemize}
    \item if $\hat{f} > T(1+\beta_2)$ then delete edge ($v_a,v_b$)
    \item else if $\hat{f} < T(1-\beta_1)$ then keep edge ($v_a,v_b$)
    \item else keep the edge with probability $\frac{1}{2}$
\end{itemize}
Update $\mathcal{G}$, $t_i=t_i+1$\\
$\epsilon_{\text{Total}}^{(i+1)}=\epsilon_{\text{Total}}^{(i)}-\left(t_i\epsilon_i^2+\epsilon_i\sqrt{2t_i\log(\frac{1}{\delta'})}\right)$
$\epsilon_{\text{curate}}^{(i)}=\left(t_i\epsilon_i^2+\epsilon_i\sqrt{2t_i\log(\frac{1}{\delta'})}\right)$\\
$\delta_\text{curate}^{(i)}=\delta'+(t_i\cdot \delta)$\\
$e_{i+1} = $  edges in updated graph $\mathcal{G}$
\If{$\sum_{j=0}^i\delta_\text{curate}^{(j)}<\delta_\text{Total}$}{Continue}}}
\Return Skeleton $\mathcal{G}$, Total Leakage ($\sum_{j=0}^{d-2}\epsilon_{\text{curate}}^{(j)},\sum_{j=0}^{d-2}\delta_{\text{curate}}^{(j)}$)
\end{algorithm}
\subsection{Adaptive Privacy Budget Allocation with Score-based \textit{CURATE} Algorithm: } \label{sec3.2}In this sub-section, we present the adaptive and non-uniform private budget allocation mechanism for the class of score-based algorithms which is based on the idea of functional causal models (FCM). Traditional score-based algorithms estimate the causal graph that optimizes a predefined score function such as Bayesian Dirichlet equivalent uniform (BDe(u)\cite{bdeu}), Bayesian Gaussian equivalent (BGe\cite{bge}), Bayesian Information Criterion (BIC \cite{BIC}), and Minimum Description Length (MDL \cite{mdl}). These methods are agnostic to the underlying true distribution of the data. There is a line of work in the literature that aims to extract more accurate underlying distributions from observational data through a functional causal model (FCM). Given an observational dataset $\mathcal{D}$ with $(\mathbf{x_1},\ldots,\mathbf{x_n})$ i.i.d. samples and $d$-number of features $\mathcal{F}=\{F_1,\ldots,F_d\}$, FCM based methods mimics the data generation process $f_i(\cdot)$ to obtain feature $F_i$ as a function of its parents ($Pa(F_i)$) and added noise $Z$ as:
\begin{align*}
    F_i=f_i(\text{Pa}(F_i))+Z.
\end{align*}
It is worth mentioning that the added noise $Z$ is independent of $Pa(F_i)$ and depends on the sensitivity of the deterministic function $f_i(\cdot)$.
As the traditional score-based algorithms impose combinatorial acyclicity constraints while learning DAG from observational data, the estimation process becomes NP-hard \cite{chickering1996learning}. To address this, the non-private FCM-based algorithm, NOTEARS \cite{zheng_dags_2018}, introduces a continuous optimization problem which optimizes score function $\mathtt{score}(W)$ as:
\begin{align}\label{eqNOTEARS}
    &\min_{W\in \mathbb{R}^{d\text{x} d}}\mathtt{score}(W)\text{    subject to } h(W)=0,
\end{align}
where the score-function $\mathtt{score}(\cdot):\mathbb{R}^{d\text{x}d}\rightarrow \mathbb{R}$ is the combination of squared loss function and a penalization function. Briefly, the score function is defined as:  
\begin{align}
\mathtt{score}(W,\alpha)&=\underbrace{\ell(W;\mathcal{D})+\lambda ||W||_1}_{\text{objective function}} +\underbrace{\frac{\rho}{2}|h(W)|^2}_{\text{quadratic penalty}} \\\nonumber &+\underbrace{\alpha h(W)}_{\text{Lagrangian multiplier}},
\end{align}
where $\rho>0$ is a penalty parameter, $\alpha$ is Lagrange multiplier, and $\lambda||W||_1$ is a non-smooth penalizing term for dense graph. The algorithm imposes the acyclicity constraint with $h:\mathbb{R}^{d\text{x} d}\rightarrow \mathbb{R}$, where $h(\cdot)$ is a smooth function over real matrices \cite{zheng_dags_2018}. The acyclicity constraint is defined by the function $h(W)$ as:
\begin{align*}
    h(W)=\text{tr}(e^{W\circ W})-d = 0,
\end{align*} 
where, $\circ$ is the \textit{Hadamard product}, and $e^{W\circ W}$ is the \textit{matrix exponential} of ${W\circ W}$. The acyclicity constraint $h(W)$ is a non-convex function and has a gradient: $\nabla h(W)=(e^{W\circ W})^\text{T}\circ 2W$ \cite{zheng_dags_2018}.
For a given dataset $\mathcal{D}\in \mathbb{R}^{n\text{x}d}$ with $n$ i.i.d. samples of feature vector $\mathcal{F}=(F_1,\ldots,F_d)$, let $\mathbb{D}$ denotes a discrete space of DAGs $\mathcal{G}=(V,E)$ on $d$ nodes. The objective of the NOTEARS algorithm \cite{zheng_dags_2018} is to model $(F_1,\ldots, F_d)$ via FCM. The $j^{th}$ feature is defined by $F_j=w_j^T\mathcal{F}+Z$ where $\mathcal{F}=(F_1,\ldots,F_d)$ is a feature vector and $Z=(z_1,\ldots,z_d)$ is a added noise vector. 
\newline

\textit{\textbf{Differentially Private score-based CGD Algorithms: }} The optimization problem mentioned in \eqref{eqNOTEARS} is non-private and therefore releasing the gradient of the optimization problem is prone to privacy leakage. To address this privacy concern, the DP-preserving score-based CGD algorithm NOLEAKS \cite{ma_noleaks_2022} adopts the notion of Differential privacy (DP) in this optimization process. To ensure differential privacy for the released gradient $(\hat{\nabla F})$, the Jacobian of this optimization process is clipped with certain clipping threshold $(s)$ and perturbed with the Gaussian noise $\mathcal{N}(0,\sigma^2\mathbf{I}_{\text{dxd}})$.
\newline
Unlike, the constraint-based CGD algorithms, later iterations are more critical compared to the initial ones in this minimization process of the score function $\mathtt{score}(W,\alpha)$. Intuitively, initial iterations of the optimization process may handle more noise but as the algorithm tends to converge to the optima, the amount of added noise needs to be reduced for better convergence. This adaptivity in terms of added noise also ensures less chances of missing the optima. Motivated by this crucial fact, we introduce adaptivity to this setting and describe our proposed framework in the next section. As the NOLEAKS algorithm perturbs the Jacobian matrix through the \textit{Gaussian noise} with \textit{same noise parameter (privacy  budget)} to achieve DP guarantee, the main difference between the existing differentially private framework NOLEAKS and our proposed framework, score-based \textit{CURATE} is the per-iteration adaptive privacy budget increment during the perturbation of the Jacobian matrix.
\newline
\textit{\textbf{Adaptive Privacy Budgeting for Score-based Algorithms: }}We observe a room for improvement in terms of adaptive privacy budget allocation for differentially private FCM-based CGD algorithms. Intuitively, the later steps/iterations in the optimization of \eqref{eqNOTEARS} are more crucial compared to the initial ones, as the later iterations are closer to the optima. Recent works including \cite{concentrated_lee},\cite{zhang_adaptive_2021}, \cite{chen2023differentially} propose several adaptive privacy budget allocation mechanisms for gradient-based optimization problem that allocate privacy budgets for each iteration adaptively in the optimization process. In our proposed framework for score-based setting, we aim to implement adaptive privacy budget allocation for each iteration and increment the privacy budget as a function of the iterations. Therefore, our goal is to select an adaptive privacy budgeting mechanism for the scope of score-based algorithms that allocates less privacy budget to the initial iterations compared to the later ones. Intuitively, privacy budgets can be incremented additively, multiplicatively, and exponentially.
Next, we analyze three different methods of incrementing the privacy budget as a function of the initial privacy budget ($\epsilon_0$), the number of iterations $(i)$ and present some experimental results to highlight the method that achieves better F1-score. 
\newline 
We analyzed the performance of three different privacy budget increment mechanisms in this paper, and next, we demonstrate the mechanisms briefly. First, we mention \textit{Additive Increment: $\epsilon_i=\epsilon_0(1+\frac{i}{I})$}. In this scheme, the privacy budget of the $i^{th}$ iteration is defined as a linear function of the initial budget ($\epsilon_0$), current iteration ($i$), and total number of iterations ($I$). 
Next, we analyze \textit{Exponential Increment: $\epsilon_i=\epsilon_0 \cdot \exp(\frac{i}{I})$}. In this scheme, the budget of the $i^{th}$ iteration gets incremented as a function of $\exp(\frac{i}{I})$. The third increment method is \textit{Multiplicative Increment: $\epsilon_i = \epsilon_0^{(1+\frac{i}{I})}$}. In this method, $\epsilon_i$ gets incremented multiplicatively as a function of $\epsilon_0^{\frac{i}{I}}$.

\begin{lm}\label{sc_curate_lemma2}
    Given a total privacy budget of $\epsilon_{\text{Total}}$, initial privacy budget $\epsilon_0$, it is possible to execute total possible number of iterations $I_{\text{add}}=\frac{\epsilon_{\text{Total}}+\frac{\epsilon_0}{2}}{\epsilon_0+\frac{\epsilon_0}{2}}$ (with additive increment), $I_{\text{exp}}=\frac{\epsilon_{\text{Total}}}{\epsilon_0\cdot \exp(1)}$ (with exponential increment) and $I_{\text{mul}}= \frac{\log(\epsilon_0)}{\log\left(1-\frac{\epsilon_0(1-\epsilon_0)}{\epsilon_{\text{Total}}}\right)}$ for $\epsilon_0<1$ and $I_{\text{mul}}= \frac{\log(\epsilon_0)}{\log\left(1+\frac{\epsilon_0(\epsilon_0-1)}{\epsilon_{\text{Total}}}\right)}$ for $\epsilon_0>1$ (with multiplicative increment).
\end{lm}
Lemma \ref{sc_curate_lemma2} shows an explicit dependence of the total number of possible iterations on the total privacy budget ($\epsilon_{\text{Total}}$) and initial privacy budget $\epsilon_0$. Figure \ref{fig_adanl_methods} shows the maximum possible number of iterations by different adaptive methods, given a fixed initial privacy budget ($\epsilon_0$) and total privacy budget ($\epsilon_{\text{Total}}$). We also observe that in high privacy regime ($\epsilon_0<1$),  the multiplicative method notably executes more number of iterations compared to the additive and exponential one. As we aim to achieve better performance by executing more number of iterations given a total privacy budget ($\epsilon_{\text{Total}}$), in this paper we follow a multiplicative method for per iteration privacy budget increment. 
\newline
\textit{\textbf{Score-based CURATE Algorithm: }} We present the adaptive private minimization technique used in score-based \textit{CURATE} in Algorithm \ref{CURATE_algo}. By using the \texttt{Priv-Linesearch} feature adopted from the algorithm NOLEAKS \cite{ma_noleaks_2022}, by which the algorithm aim to investigate the optimal step size $\eta$. The score-based \textit{CURATE} algorithm essentially utilizes the FCM-based models for CGD and allows adaptive privacy budgeting through the optimization process.
Score-based \textit{CURATE} algorithm follows a similar FCM-based framework as the non-private NOTEARS algorithm and differentially private NOLEAKS algorithm, however our proposed framework allows adaptive privacy budget allocation for each iteration through \texttt{Adaptive Priv-Minimize} function.
\begin{algorithm}[H]
\caption{\texttt{Adaptive Priv-Minimize}}
 \SetKwInput{KwData}{Input}
 \SetKwInput{KwResult}{Output}
 \KwData{$\nabla F$: Gradient of the objective function, $W_0$: Initial Guess, $\epsilon_0$: Initial privacy budget, $\delta$: Failure probability in DP}
 \KwResult{$W$: Adjacency Matrix}
 compute noise parameter $\sigma$ according to Equation \eqref{sigma}\\
 $\hat{\nabla F}\leftarrow clip(\nabla F|_{W=W_0})+\mathcal{N}(0,\sigma^2\textbf{I}_{d\text{x}d})$\\
 $\forall i=j,\hat{\nabla F_{0ij}}\leftarrow 0$
\\
\For{$k = 0,\ldots, I-1$}
{$\epsilon_k=\epsilon_0^{(1+\frac{k}{I-1})}$\\
recompute noise parameter $\sigma$ according to Equation \eqref{sigma} with $\epsilon_k$ and $\delta$\\
compute the direction $p_k$ with the clipped gradient $\hat{\nabla F_k}$\\
$\eta \leftarrow \texttt{Private-LineSearch()}$\quad {[decide the step size]}\\
$s_k\leftarrow \eta p_k$\\
$W_{k+1}\leftarrow W_k + s_k$\\
\If{$k<\text{I}-1$}
{$\hat{\nabla F_{k+1}}\leftarrow \mathtt{clip}(\nabla F|_{W=W_0})+\mathcal{N}(0,\sigma^2\textbf{I}_{d\text{x}d})$;\\
$\forall i=j, \hat{\nabla F_{{k+1}ij}}\leftarrow 0$;\\
update auxiliary data;}
}
\Return{Adjecency matrix after $I^{th}$ iteration: $W_I$ }
\label{CURATE_algo}
\end{algorithm}

\textit{\textbf{Remarks on score-based CURATE Algorithm}: } As the score-based \textit{CURATE} algorithm follows a similar FCM based workflow as the non-private NOTEARS and differentially private NOLEAKS algorithm, it achieves polynomial complexity in terms of the feature/variable size $d$. For small datasets and with less leakage, it achieves better and more meaningful causal graphs compared to the constraint-based algorithms. However, due to the non-convex nature of the optimization problem, similar to NOTEARS and NOLEAKS algorithms, score-based \textit{CURATE} algorithm does not guarantee convergence to global optima.
Nonetheless, experimentally we observe that score-based algorithms ensure better privacy guarantees in lower total privacy regime ($\epsilon_{\text{Total}}\le 1$) compared to the differentially private constraint-based algorithms.
\begin{figure*}[t]
    \centering
    \includegraphics[scale=0.18]{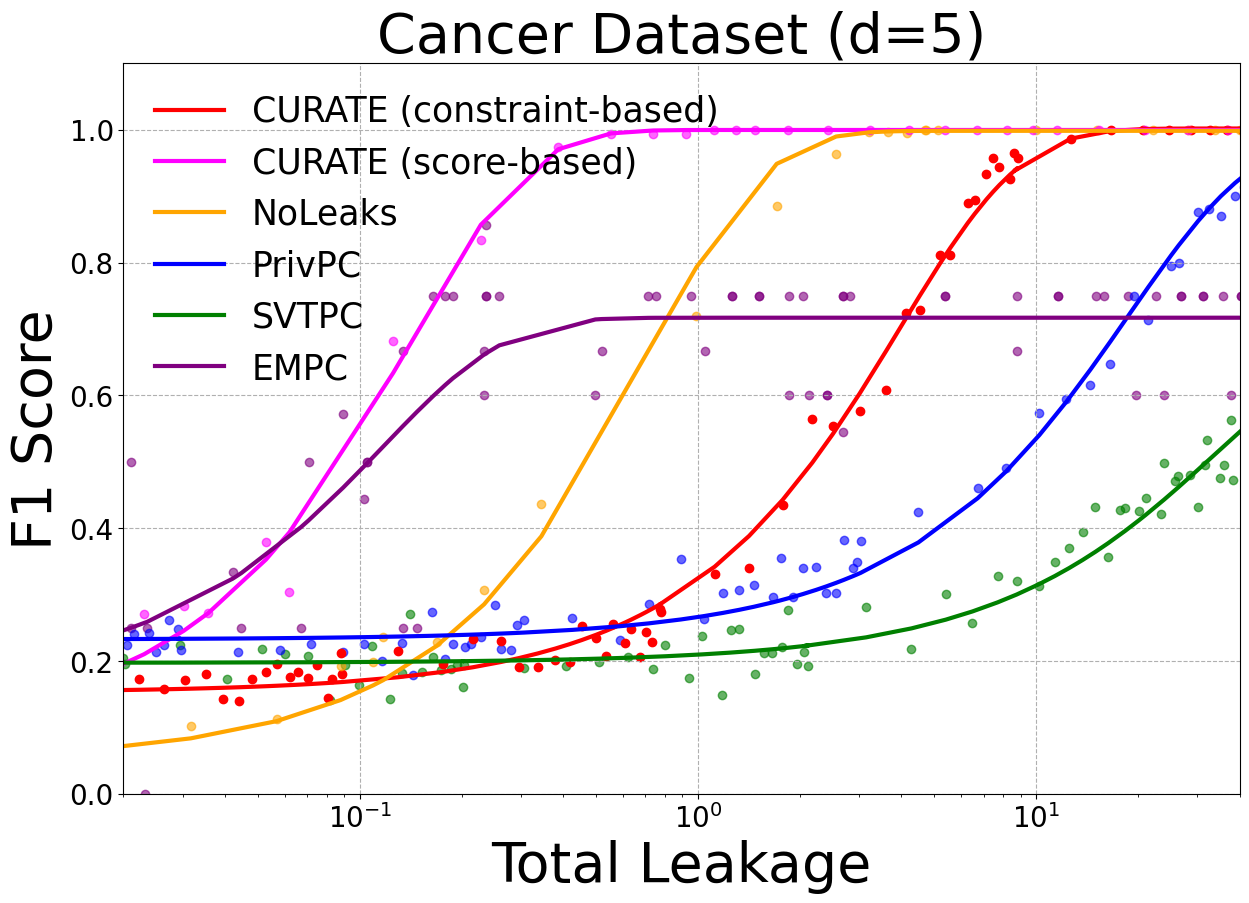}
    \includegraphics[scale=0.18]{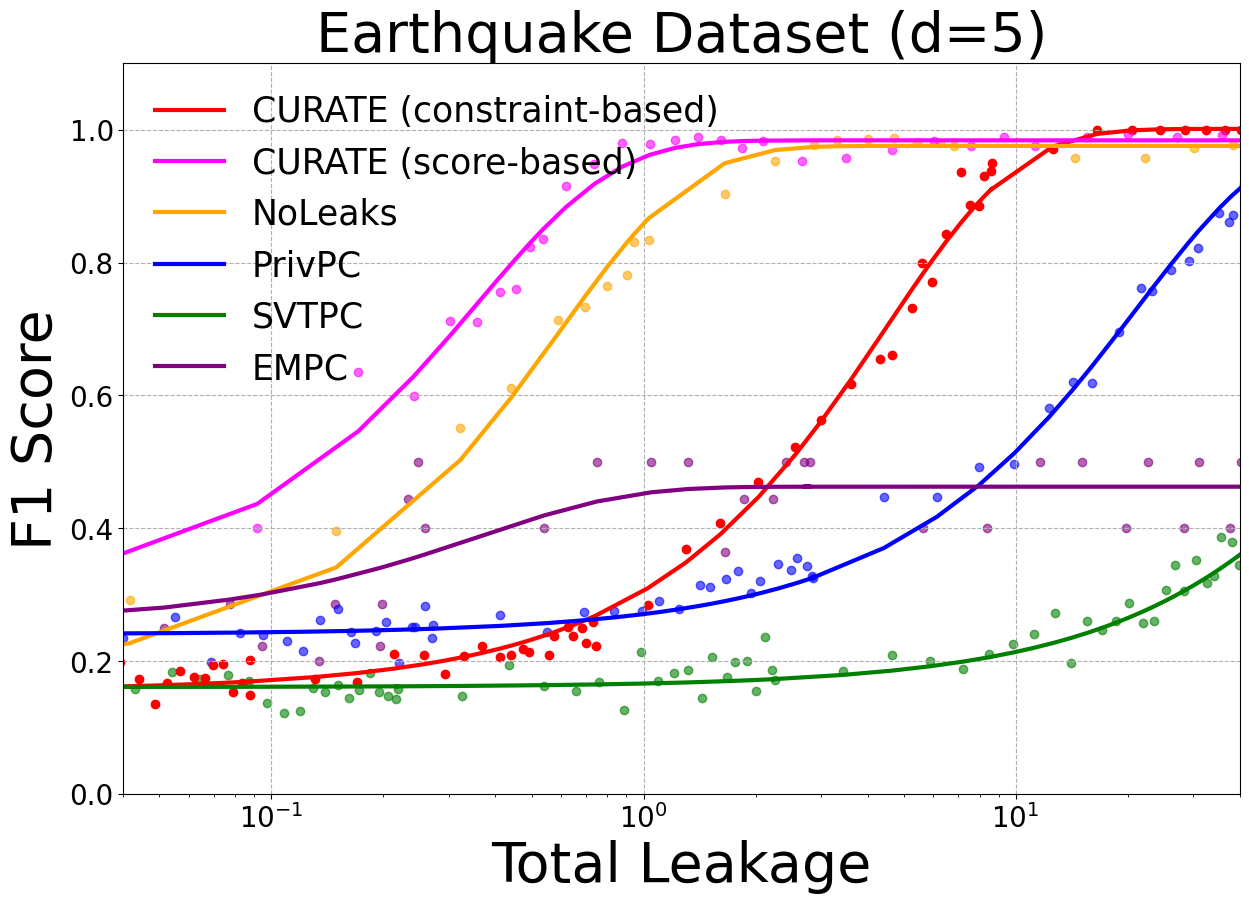}
    \includegraphics[scale=0.18]{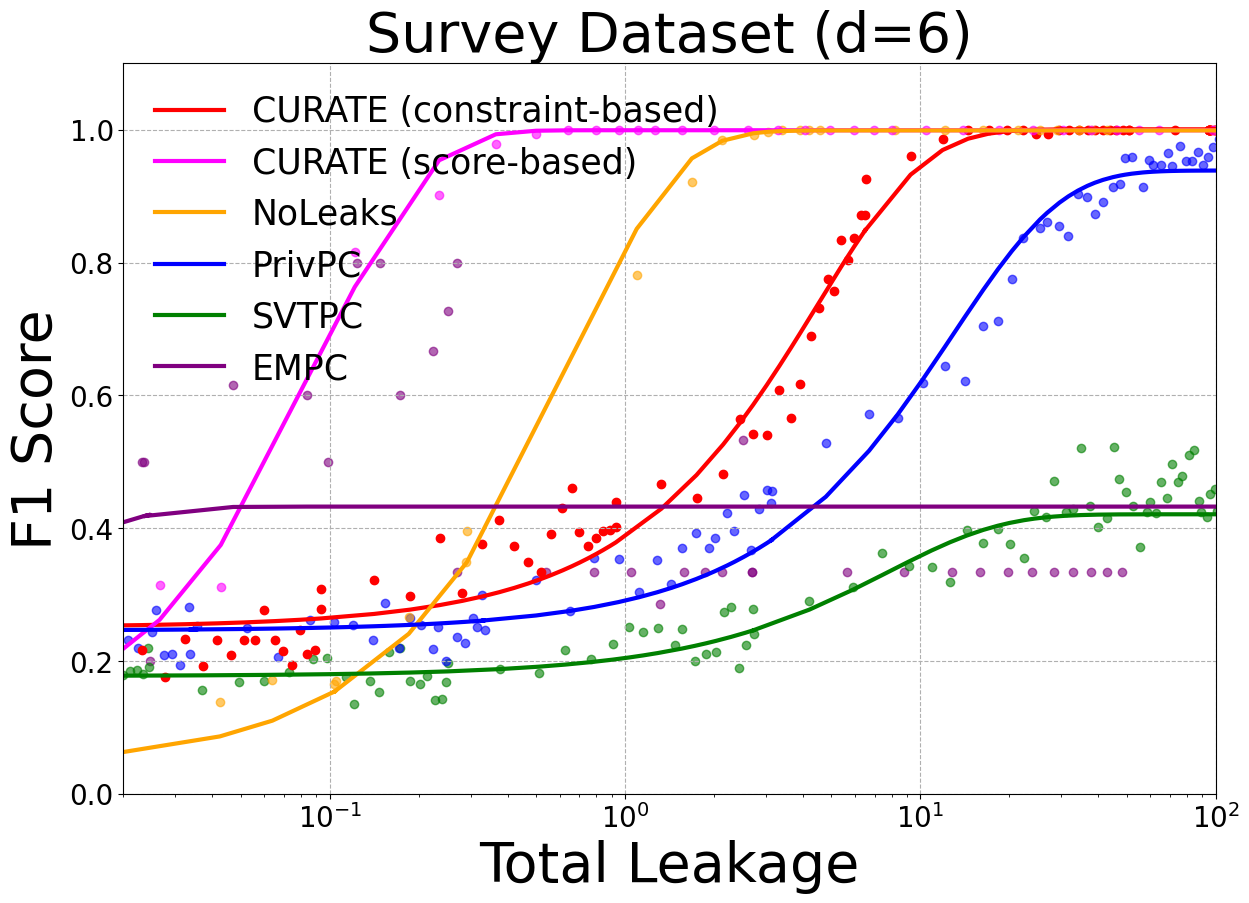}
    \includegraphics[scale=0.18]{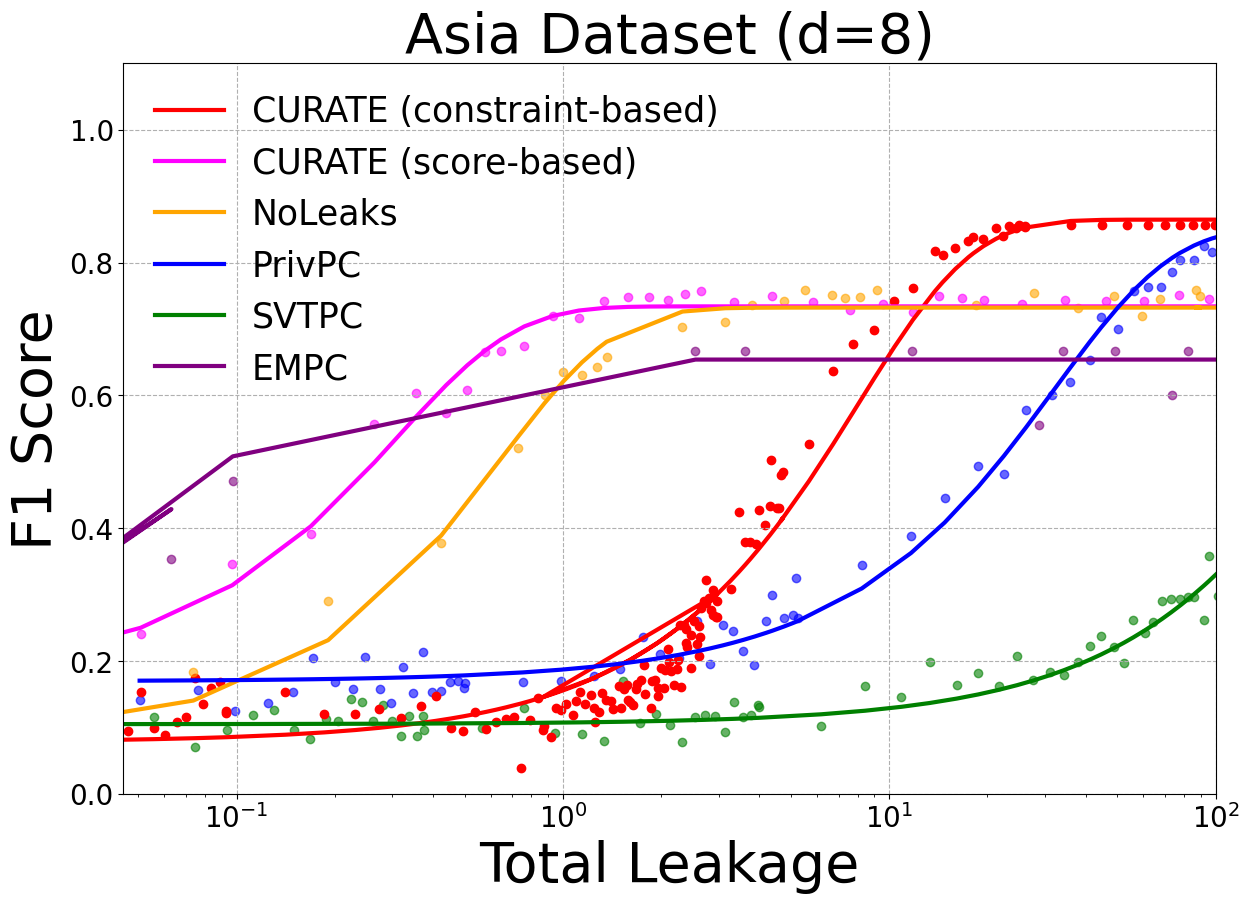}
    \includegraphics[scale=0.18]{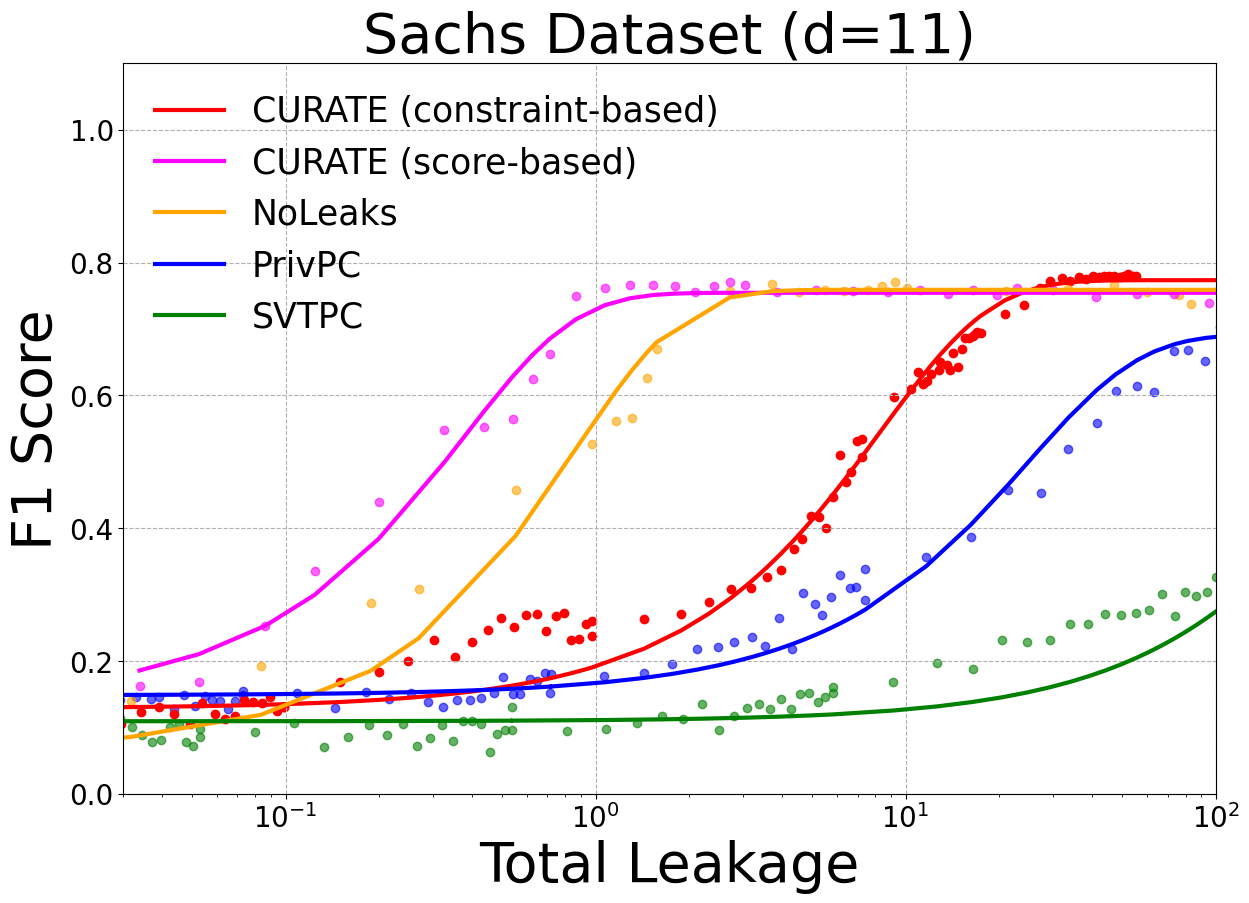}
    \includegraphics[scale=0.18]{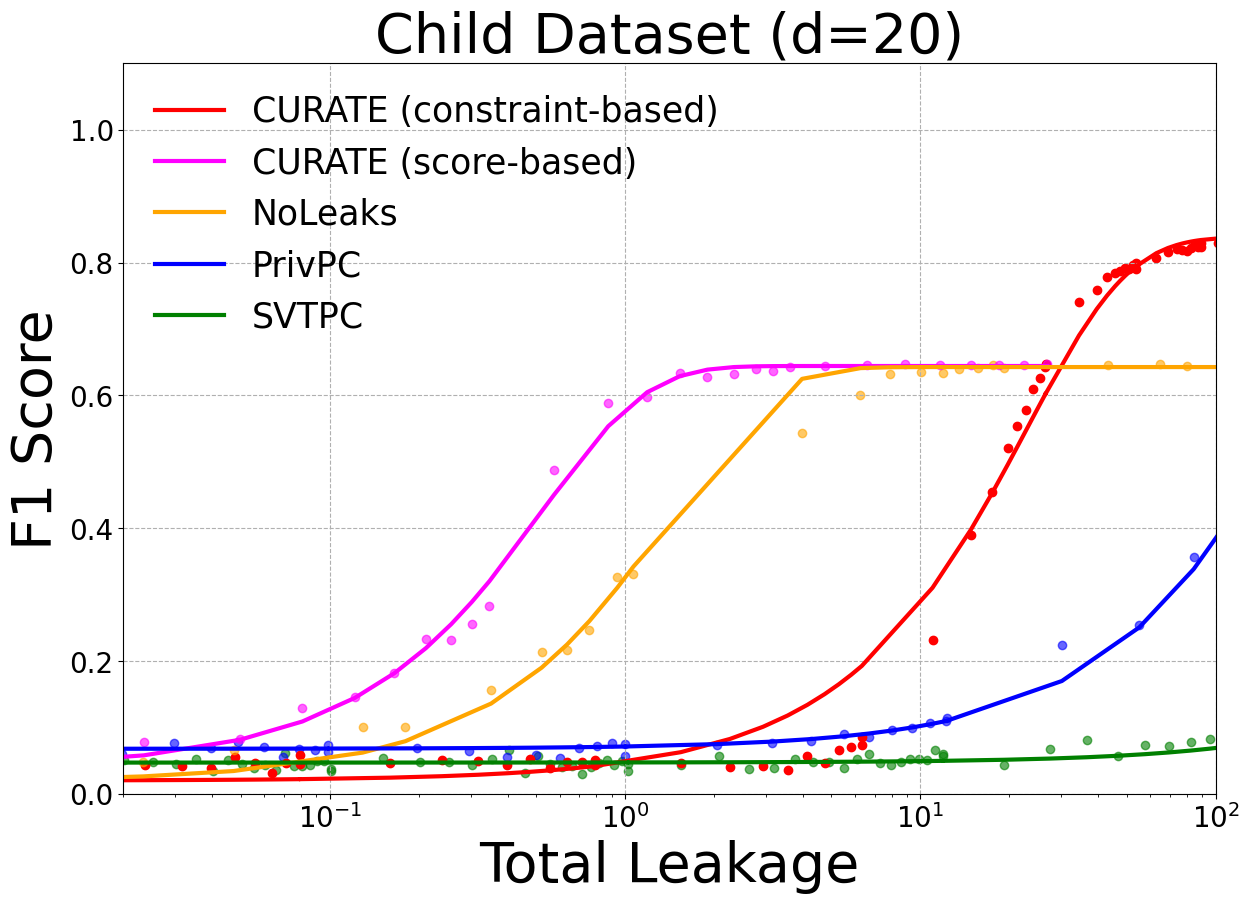}
    \includegraphics[scale = 0.27]{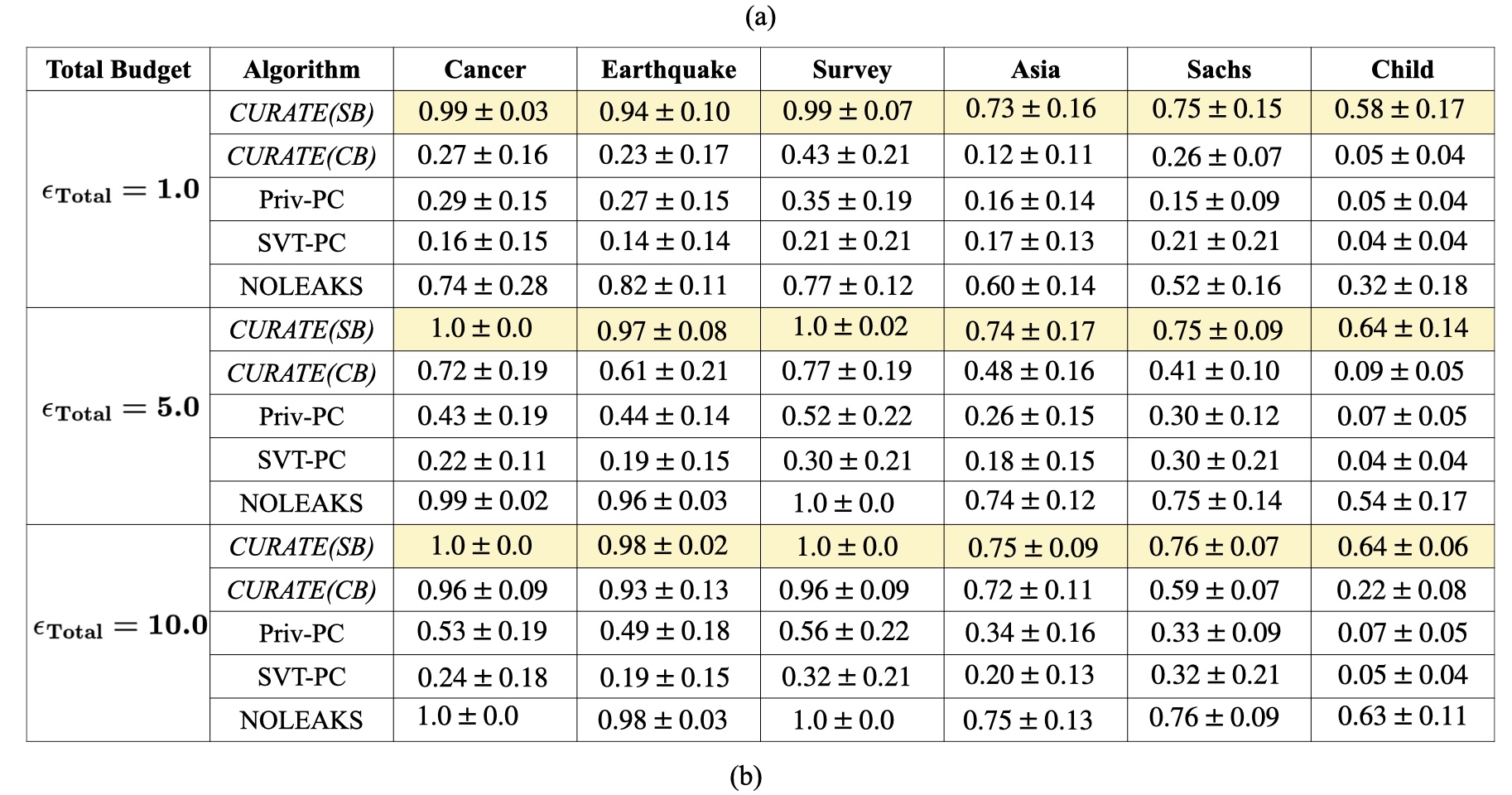}
    \caption{Part (a) represents the performance evaluation of differentially private CGD algorithms EM-PC \cite{xu_differential_2017}, SVT-PC, Priv-PC \cite{wang_towards_2020}, NOLEAKS \cite{ma_noleaks_2022} and \textit{CURATE} (score-based and constraint-based) in terms of total leakage vs F1 score on 6 public CGD datasets: Cancer, Earthquake, Survey, Asia, Sachs, Child. Part (b) presents the mean and standard deviation of F1-score for 50 consecutive runs for three privacy regimes ($\epsilon_{\text{Total}} = 0.1$, $\epsilon_{\text{Total}}=5.0$, $\epsilon_{\text{Total}}=10.0$).}
    \label{TLvsF1}
\end{figure*}\begin{figure*}[t]
    \centering
    \includegraphics[scale=0.27]{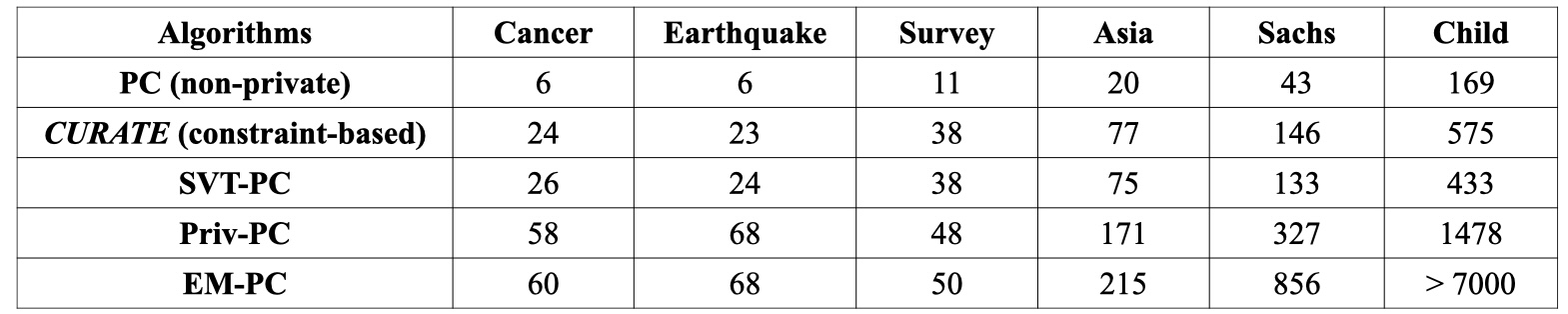}
    \caption{Average CI tests required to achieve the maximum F1 score with comparatively large amount of total leakage ($\epsilon_{\text{Total}}=1.0$) on Cancer, Earthquake, Survey, Asia, Sachs, and Child datasets. Average CI tests in \textit{CURATE} converge to the non-private PC algorithm whereas EM-PC \cite{zheng_dags_2018}, Priv-PC and SVT-PC  \cite{wang_towards_2020} tend to run more CI tests.}
    \label{table-CItests}
\end{figure*}
\section{Results and Discussion: }
\label{sec:results}
\begin{figure*}[t]
    \centering
    \includegraphics[scale=0.27]{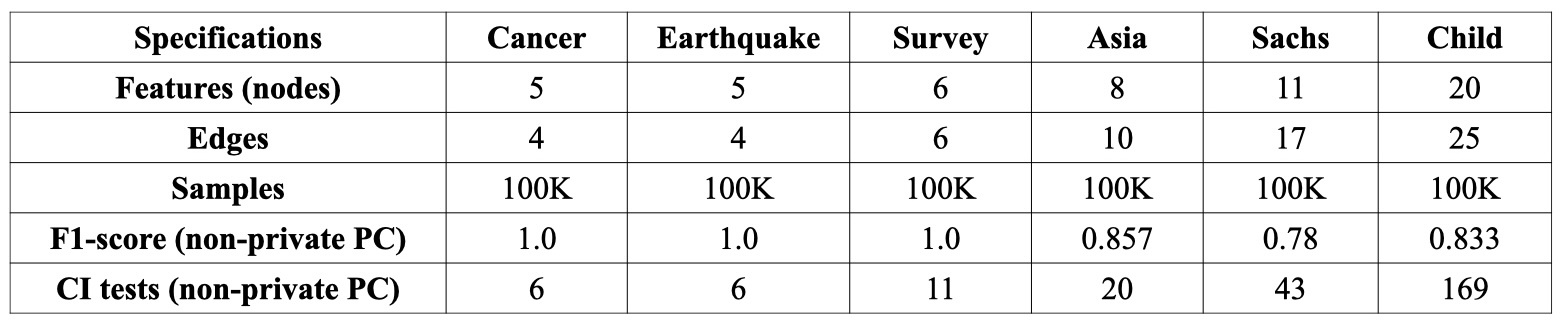}
    \caption{Dataset description and CGD results of non-private PC algorithm \cite{spirtes_causation_1993} on 6 public CGD datasets with Kendall's $\tau$ CI test statistic (The results are obtained with the following parameters: sub-sampling rate $=1.0$, test threshold $= 0.05)$.}
    \label{data_description}
\end{figure*}
\begin{figure*}[t]
    \centering
    \includegraphics[scale=0.27]{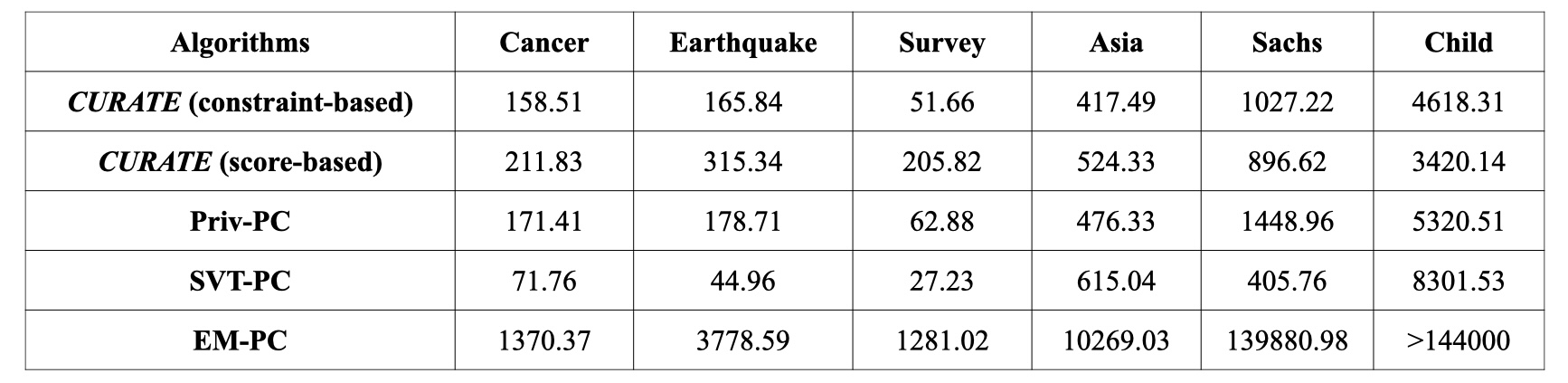}
    \caption{Running time comparison of differentially private constraint-based and score-based algorithms on 6 public CGD datasets: Cancer, Earthquake, Survey, Asia, Sachs, and Child (in seconds) for 50 consecutive iterations.}
    \label{runtime}
    \vspace{-1em}
\end{figure*}
\textit{\textbf{Data Description and Test Parameters: }}We compared the predictive performance of our proposed framework \textit{CURATE} with non-private PC \cite{spirtes_causation_1993}, EM-PC \cite{xu_differential_2017}, SVT-PC, Priv-PC \cite{wang_towards_2020} and NOLEAKS \cite{ma_noleaks_2022} on 6 public CGD datasets \cite{korb_bayesian_nodate,bernardo_bayesian_1992,lauritzen_local_1988,sachs_causal_2005,denis_bayesian_2014}. Table \ref{data_description} presents the detailed description of the datasets along with the predictive performance of the non-private PC algorithm.
For the experimental results, we considered the probability of failure in differential privacy ($\delta'=10^{-12}$), as the safe choice for $\delta'$ is ($\delta'\le n^{-1.5}$) where $n$ is the total number of participants/samples in the dataset. In each of the 6 CGD datasets, the total number of samples $(n)= 100k= 10^5$, and thus we considered the value of $\delta'= 10^{-12}\le n^{-1.5}$. The test threshold $(T)$ is set as $0.05$, sub-sampling rate $(q)$ as $1.0$, and we have used Kendall's $ \tau$ as a CI testing function for the constraint-based private algorithms. To run the experiments \footnote{The code for constraint-based and score-based CURATE algorithm is available at: https://github.com/PayelBhattacharjee14/cgdCURATE}, we have used a high-performance computing (HPC) system with 1 node and 1 CPU with 5GB RAM. 
\newline
\textit{\textbf{Evaluation Metric: }} For the scope of our experiments, we measured the predictive performance of a CGD algorithm in terms of F1-score which indicates the similarity between the estimated graph ($\mathcal{G}$) and the ground truth ($\mathcal{G}^*$).
Let the ground truth is represented by the graph $\mathcal{G}^*=(\mathcal{V},\mathcal{E}^*)$ and the estimated graph is represented by $\mathcal{G}=(\mathcal{V},\mathcal{E})$.
Then by defining Precision= $\frac{\mathcal{E}\cap\mathcal{E}^*}{\mathcal{E}}$, and Recall= $\frac{\mathcal{E}\cap\mathcal{E}^*}{\mathcal{E}^*}$, the F1-score (utility) of the CGD algorithm can be defined as:
\begin{equation*}
    F1= \frac{2\cdot \text{Precision}\cdot\text{Recall}}{\text{Precision}+\text{Recall}}.
\end{equation*}

\textit{\textbf{Privacy vs Utility Trade-off: } }There is a privacy-utility trade-off in differential privacy-preserving CGD. Through comprehensive experimental results on 6 public CGD datasets, we observed that the private algorithms require higher privacy leakage to achieve the same predictive performance as their non-private counterparts. The experimental result presented in Figure \ref{TLvsF1} shows that with adaptive privacy budget allocation and minimization of total probability of error, \textit{CURATE} outperforms the existing private CGD algorithms including EM-PC \cite{xu_differential_2017}, SVT-PC, Priv-PC \cite{wang_towards_2020}, and NOLEAKS \cite{ma_noleaks_2022}.
In Table  Table \ref{TLvsF1} we present the mean F1-score and its standard deviation for 50 consecutive runs on the Cancer, Earthquake, Survey, Asia, Sachs, and Child datasets for different privacy regimes.
The number of features in the dataset also impacts the performance of the CGD algorithms. Notably, for Cancer, Earthquake, and Survey datasets, score-based \textit{CURATE} achieves the highest F1-score with a total leakage of less than 1.0. But as the number of features increases, \textit{CURATE} and the other CGD algorithms tend to leak more in order to achieve the best F1-score. For Sachs and Child datasets, \textit{CURATE} achieves the highest F1-score with ($\epsilon_{\text{Total}} > 1.0$). We also observe that constraint- based \textit{CURATE} achieves better utility (F1-score) with less amount of total leakage compared to the existing constraint-based DP-CGD algorithms including EM-PC \cite{zheng_dags_2018}, Priv-PC, SVT-PC  \cite{wang_towards_2020}. Therefore the adaptive privacy budgeting scales-up utility in DP-CGD.
\newline
\textit{\textbf{Computational Complexity of DP-CGD Algorithms: }}The reliability of an algorithm also depends on the computational complexity. In private CGD, score-based and constraint-based algorithms have different computational complexities. As mentioned by the authors \cite{nogueira_methods_2022}, score-based algorithms are computationally expensive as they enumerate and assign scores to each possible output graph. For instance, NOLEAKS uses \textit{quasi-Newton} method which has high computational and space complexity \cite{ma_noleaks_2022}. On the other hand, EM-PC is computationally slow as the utility function used in \textit{Exponential Mechanism} is computationally expensive \cite{wang_towards_2020}. Priv-PC adopts SVT and \textit{Laplace Mechanism} to ensure DP whereas, constraint-based \textit{CURATE} optimizes privacy budgets ($\Bar{\epsilon}$) in an online setting and then adopts \textit{Laplace Mechanism} to privatize CI tests. This makes \textit{CURATE} computationally less expensive compared to the existing constraint-based  DP-CGD algorithms.
\newline
\noindent \textbf{\textit{Comparison of Number of CI Tests: }} The total number of CI tests executed by a differentially private CDG algorithm directly affects the privacy and utility trade-off of the algorithm. The total number of CI tests in private constraint-based CGD algorithms directly influences the total amount of leakage as each CI test is associated with some amount of privacy leakage. The privacy leakage can be provably reduced by efficient and accurate CI testing. In the constraint-based \textit{CURATE} algorithm, the privacy budgets are allocated by minimizing the surrogate for the total probability of error. Intuitively, in \textit{CURATE}, the total leakage decreases as the adaptive choice of privacy budgets makes the initial CI tests more accurate, and therefore,  \textit{CURATE} tends to run a smaller number of CI tests compared to other differentially private algorithms. We confirm this intuition in the results presented in Table \ref{table-CItests}. We observe that the number of CI tests in EM-PC, SVT-PC, and Priv-PC are comparatively large to \textit{CURATE} and the non-private counterpart PC algorithm \cite{spirtes_causation_1993}. 
\newline
\noindent\textit{\textbf{Running Time Comparison: }} In this subsection of the paper, we address the run-time comparison between adaptive and non-adaptive score-based and constraint-based differentially private CGD algorithms. Due to the complexity of the algorithms, score-based CGD algorithms tend to consume more time compared to constraint-based algorithms. In Table \ref{runtime}, we compare the run-time of the existing differentially private CGD algorithms for 50 consecutive iterations. As presented in Table \ref{runtime}, the constraint-based \textit{CURATE} algorithm speeds up the process of DP-CGD compared to Priv-PC and EM-PC algorithms. The score-based \textit{CURATE} algorithm achieves better predictive performance compared to the NOLEAKS algorithm with similar amount of execution time. Therefore, we can observe that adaptivity enables the DP-CGD algorithms to converge faster and reduces the overall execution time.
\section{Conclusion}
In this paper, we propose a differentially private causal graph discovery framework \textit{CURATE} that scales up privacy by adaptive privacy budget allocation for both constraint-based and score-based CGD environment. Constraint-based \textit{CURATE} is based on the key idea of minimizing the total probability during adaptive privacy budgeting, and this ensures a better privacy-utility trade-off. The score-based \textit{CURATE} framework allows higher number of iterations and faster convergence of the optimization problem by adaptive budgeting, hence it guarantees better utility with less leakage. We observe that the average required CI tests in constraint-based \textit{CURATE} is compared to the existing DP-CGD algorithms and it is close to the number of CI tests of the non-private PC algorithm. Experimental results show that \textit{CURATE} outperforms the existing private CGD algorithms and achieves better utility with leakage smaller by orders of magnitude through adaptive privacy budgeting. There are several interesting open research directions for future work: (i) implantation of adaptive gradient-clipping mechanism for the score-based DP-CGD algorithms, (ii) our proposed framework uses the resulting prune graph, the per iteration privacy budget can be designed based on the outcome of the previous iteration for score-based algorithms, (iii) the outcomes of the previous noisy test can be used to tune the hyper-parameters including test threshold, margins, and clipping thresholds.

\section{Appendix}
\subsection{Proof of Lemma \ref{cb_curate_lemma1}}\label{appA}
In this Section, we present the proof of Lemma $1$. For every order-$i$ conditional independence (CI) test, we have a privacy budget of $\epsilon_i$. 
Given a CI test statistic $f(\mathcal{D})$ with $l_1$-sensitivity $\Delta_1$, threshold $T$ and margins $(\beta_1,\beta_2)$, we perturb the test statistic by \textit{Laplace noise} defined as $Z=Lap(\frac{\Delta_1}{\epsilon_i})$, and check for conditional independence between $(v_a,v_b)\in \mathcal{G}$ conditioned on $S$ as:
\begin{enumerate}
    \item If $f(\mathcal{D})+Z > T(1+\beta_2)\implies$ \text{delete edge} $(v_a,v_b)$,
    \item If $f(\mathcal{D})+Z < T(1-\beta_1)\implies$ keep edge $(v_a,v_b)$,
    \item Else keep edge ($v_a,v_b$) with probability $\frac{1}{2}$. 
\end{enumerate}
For simplicity of notations, we define $f_{v_a,v_b|S}(\mathcal{D}):=f(\mathcal{D})$.
\newline 
\textbf{Type-I Error: }
We now analyze the Type-I error relative to the unperturbed CI test, i.e., the private algorithm keeps the edge given that the unperturbed test statistic deletes the edge $f(\mathcal{D})> T$. In other words, this can be written as: $\mathbb{P}(E_{1}^{i})= \mathbb{P}(\text{Error}|f(\mathcal{D})> T)$.
We next note that the error event occurs only for cases (b) and (c). We can bound the relative Type-I error as follows:
\begin{align}
    &\mathbb{P}(E_{1}^{i})= \mathbb{P}(\text{Error}|f(\mathcal{D})> T)\\
    &\leq \frac{1}{2}\left(\mathbb{P}(f(\mathcal{D}) + Z\in [T(1-\beta_1), T(1+\beta_2)]| f(\mathcal{D})>T)\right) \nonumber \\\nonumber
    & \quad +  \mathbb{P}(f(\mathcal{D}) + Z< T(1-\beta_1)| f(\mathcal{D})>T)\\\nonumber
&\leq \frac{c_1}{2} + \mathbb{P}(f(\mathcal{D}) + Z< T(1-\beta_1)| f(\mathcal{D})>T)\\\nonumber
&\leq \frac{c_1}{2} + \frac{1}{2}\exp\left(\frac{-T\beta_1\epsilon_i}{\Delta_1}\right),\\
\end{align}
where the last inequality follows from the Laplacian tail bound and using the fact that $f(\mathcal{D})>T$; and we have defined $c_1$ as 
$c_1 := \mathbb{P}(f(\mathcal{D}) + Z\in [T(1-\beta_1), T(1+\beta_2)]| f(\mathcal{D})>T)$. 
\newline
Upper-bound on $\mathbb{P}[f(\mathcal{D})+Z<T(1-\beta_1)|f(\mathcal{D})>T]$ is obtained from \textit{Laplace Tail bound} as:
\begin{align}
    &\mathbb{P}[f(\mathcal{D})+Z<T(1-\beta_1)|f(\mathcal{D})>T]\\\nonumber
    &=\mathbb{P}[Z<T(1-\beta_1)-f(\mathcal{D})]\nonumber\\
    &=\frac{1}{2}\exp\left(\frac{T-T\beta_1-f(\mathcal{D})}{\Delta_1/\epsilon_i}\right)\le \frac{1}{2}\exp\left(\frac{T-T\beta_1-T}{\Delta_1/\epsilon_i}\right)\nonumber\\
    &= \frac{1}{2}\exp\left(\frac{-T\beta_1\epsilon_i}{\Delta_1}\right).
\end{align}
\textbf{Type-II Error: }
Next, we analyze the Type-II error relative to the unperturbed CI test, i.e., the differentially private algorithm deletes an edge given that the unperturbed CI test statistic keeps the edge, $f(\mathcal{D})<T$. Mathematically, 
$  \mathbb{P}[E_2^i]=\mathbb{P}(\text{Error}|f(\mathcal{D})<T).$
The type-II error occurs only for cases (a) and (c). Therefore, we can bound the Type-II error as:
\begin{align}
    &\mathbb{P}(E_{2}^{i})= \mathbb{P}(\text{Error}|f(\mathcal{D})<T)\nonumber\\
    &\leq \frac{1}{2}\left(P(f(\mathcal{D}) + Z\in [T(1-\beta_1), T(1+\beta_2)]| f(\mathcal{D})<T)\right) \nonumber \\\nonumber
    & \quad +  \mathbb{P}(f(\mathcal{D}) + Z> T(1+\beta_2)| f(\mathcal{D})<T)\\\nonumber
&\leq \frac{c_2}{2} + \mathbb{P}(f(\mathcal{D}) + Z> T(1+\beta_2)| f(\mathcal{D})<T)\\\nonumber
&\leq \frac{c_2}{2} + \frac{1}{2}\exp\left(\frac{-T\beta_2\epsilon_i}{\Delta_1}\right),\\
\end{align}\label{typeIIdeledge}
where the last inequality follows from the Laplacian tail bound and using the fact that $f(\mathcal{D})<T$; and we have defined $c_2$ as 
$c_2 := \mathbb{P}(f(\mathcal{D}) + Z\in [T(1-\beta_1), T(1+\beta_2)]| f(\mathcal{D})<T)$.
The probability $\mathbb{P}[f(\mathcal{D})+Z>T(1+\beta_2)|f(\mathcal{D})<T]$ can also be upper bounded as:
\begin{align}
    \mathbb{P}[f(\mathcal{D})&+Z>T(1+\beta_2)|f(\mathcal{D})<T]\nonumber\\
    &=\mathbb{P}[Z>T(1+\beta_2)-
    f(\mathcal{D})]\nonumber\\\nonumber
    &=\frac{1}{2}\exp\left(-\frac{T+T\beta_2-f(\mathcal{D})}{\Delta_1/\epsilon_i}\right)\\\nonumber
    &\le \frac{1}{2}\exp\left(-\frac{T+T\beta_2-T}{\Delta_1/\epsilon_i}\right)\\\nonumber
    &= \frac{1}{2}\exp\left(\frac{-T\beta_2\epsilon_i}{\Delta_1}\right).\\
\end{align}
This concludes the proof of Lemma \ref{lemma1}.
\vspace{-1em}
\subsection{Sensitivity Analysis of Weighted Kendall's $\tau$: }\label{appB}
Conditional independence (CI) tests in Causal Graph Discovery (CGD) measure the dependence of one variable ($v_a$) on another ($v_b$) conditioned on a set of variables.
Let, the CI test statistic for connected variable pairs ($v_a,v_b$) in graph $\mathcal{G}$ is ${\tau(\mathcal{D})}$ for dataset $\mathcal{D}$ and ${\tau(\mathcal{D}')}$ for dataset $\mathcal{D}'$. For large samples, the test statistic $\tau(\cdot)$ follows a \textit{Gaussian Distribution}. Therefore, the sensitivity of the can be defined as: 
\begin{align}
    &\Delta_1(\Phi({\tau(\mathcal{D})}))=\sup_{\mathcal{D}\neq \mathcal{D}'}|\Phi({\tau(\mathcal{D})})-\Phi({\tau(\mathcal{D}')})|\nonumber\\\nonumber
    &\le \Delta(\Phi(\cdot))\cdot \Delta({\tau(\cdot)})\nonumber\\
    &=\sup_{\mathcal{D}\neq \mathcal{D}'}\frac{|\Phi({\tau(\mathcal{D})})-\Phi({\tau(\mathcal{D}')})|}{|{\tau(\mathcal{D})}-{\tau(\mathcal{D}')}|}\cdot|{\tau(\mathcal{D})}-{\tau(\mathcal{D}')}|\nonumber\\
    &\le L_{\Phi}\cdot \sup|{\tau(\mathcal{D})}-{\tau(\mathcal{D}')}|.
\end{align}
Here, $\sup_{\mathcal{D}\neq \mathcal{D}'}|\tau{(\mathcal{D})}-\tau{(\mathcal{D}')}|$ is the $l_1$-sensitivity of the CI test statistic for dataset $\mathcal{D}$ and $\mathcal{D}'$, and $\Phi$ is the PDF of standard normal distribution. As, $\Phi(\cdot)$ is differentiable, therefore the Lipschitz constant ($\text{L}_{\Phi}$) can be upper bounded as $\text{L}_{\Phi}\le \frac{1}{\sqrt{2\pi}}$. Therefore, the sensitivity can easily be calculated with the sensitivity of the weighted test statistic.
\newline
\textbf{$l_1$-sensitivity analysis: }
For large sample size ($n>>1$), Kendall's $\tau$ test statistic follows Gaussian Distribution with zero mean and variance $\frac{2(2n+5)}{9n(n-1)}$ where $n$ is the number of i.i.d. samples. Given a dataset $\mathcal{D}$ with $d$-features, the conditional dependence of between variables ($v_a,v_b$) conditioned on set $S$ can be measured with Kendall's $\tau$ as a CI test statistic. For instance, the data is split according to the unique values of set $S$ into $k$-bins. For each $i^{th}$-bin test statistic $\tau_i$ is calculated and the weighted average of all $\tau_i$ represents the test statistic for the entire dataset. The weighted average[8] is defined as:
$\tau=\frac{\sum_{i=1}^kw_i\tau_i}{\sqrt{\sum_{i=1}^kw_i}}$,
where $w_i$ is the inverse of the variance $w_i=\frac{9n_i(n_i-1)}{2(2n_i+5)}$.
\newline
\noindent As we perturb the \textit{p-value} obtained from this weighted test statistic, we need to observe the $l_1$-sensitivity of \textit{p-value}. For the scope of this paper, we consider the \textit{Lipschitz Constant} of Gaussian distribution while calculating the sensitivity.
\newline
The weighted average ($\tau$) essentially follows the standard normal distribution, i.e., $\tau\sim\mathcal{N}(0,1)$. Hence, the $l_1$-sensitivity of \textit{p-value} can be defined as:
\begin{align}
    \Delta_1&= |\Phi(\tau(\mathcal{D})-\Phi(\tau(\mathcal{D}')|\nonumber\\
    &=\frac{|\Phi(\tau(\mathcal{D})-\Phi(\tau(\mathcal{D}')|}{|\tau(\mathcal{D})-\tau(\mathcal{D}')|}\cdot |\tau(\mathcal{D})-\tau(\mathcal{D}')|\nonumber\\
    &\le \text{L}_{\Phi}|\tau(\mathcal{D})-\tau(\mathcal{D}')|\le \frac{1}{\sqrt{2\pi}}|\tau(\mathcal{D})-\tau(\mathcal{D}')|.
\end{align}\label{delta}
The sensitivity of weighted Kendall's $\tau$ can be expressed as:
\begin{align*}
    \Delta_1(\tau)&=\max_{|\mathcal{D}'-\mathcal{D}|\le 1}|\tau(\mathcal{D}')-\tau(\mathcal{D})|\le \Delta_1(\tau_i)\Delta_1(w_i).
\end{align*}
The sensitivity of $\tau_i$ depends upon the number of elements $n_i$ and $\Delta_1(\tau_i)\le \frac{2}{n_i-1}$[8]. The sensitivity of weights $\Delta(w_i)$ can be represented as follows:
\begin{align}
    \Delta_1(w_i)&\le \left\lvert\frac{w_i'}{\sqrt{\sum_{i\neq j}^k w_j+w_i'}}-\frac{w_i}{\sqrt{\sum_{j=1}^k w_j}}\right\rvert \nonumber\\
    &\le\left\lvert\frac{\frac{9n_i(n_i+1)}{2(2(n_i+1)+5)}}{\sqrt{\sum_{j=1}^kw_j+w_i'}}\right\rvert-\left\lvert\frac{\frac{9n_i(n_i-1)}{2(2n_i+5)}}{\sqrt{\sum_{j=1}^k}w_j}\right\rvert.
\end{align}\label{deltawi}
Through triangle inequality, we can provide an upper-bound on Equation \eqref{deltawi} and the sensitivity of the weight can be bounded as:
\begin{align}
    \Delta_1(w_i)\le\sqrt{\frac{2}{n}}\left(\left\lvert\frac{9n_i(n_i+1)}{2(2(n_i+1)+5)}\right\rvert-\left\lvert\frac{9n_i^2}{2(2n_i+5)}\right\rvert\right).
\end{align}\label{delwi}
The sensitivity $\Delta(\tau)$ essentially depends upon the number of elements in the $i^{th}$ bin (the bin that changed due to the addition or removal of a single user). For a dataset with block size at least size c and $kc\approx n$, with Equation \eqref{delta} and Equation \eqref{delwi}, the overall sensitivity for the p-value can be bounded as:
\begin{align}
    \Delta_1 &\le \frac{1}{\sqrt{2\pi}}\cdot\frac{2}{n_i-1}\cdot \sqrt{\frac{2}{n}}\cdot \nonumber
    \\&\left({\left\lvert\frac{9n_i(n_i+1)}{2(2(n_i+1)+5)}\right\rvert-\left\lvert\frac{9n_i^2}{2(2n_i+5)}\right\rvert}\right)
    \nonumber\\
    &=\frac{2}{\sqrt{n\pi}}\left(\frac{\left\lvert\frac{9n_i(n_i+1)}{2(2(n_i+1)+5)}\right\rvert-\left\lvert\frac{9n_i^2}{2(2n_i+5)}\right\rvert}{n_i-1}\right)
\end{align}
This concludes the $l_1$-sensitivity analysis of the weighted Kendall's $\tau$ coefficient.
\vspace{-1em}
\subsection{Proof of Lemma \ref{sc_curate_lemma2}}
Now to analyze the methods adopted for the class of score-based algorithms, we present the proof of Lemma \ref{sc_curate_lemma2}. The main objective is to derive the relationship between total privacy leakage ($\epsilon_{\text{Total}}$), number of iterations ($I$), and the initial privacy budget ($\epsilon_0$) for \textit{CURATE} (score-based) algorithm. Now, we demonstrate the possible number of iterations for Additive, Multiplicative, and Exponential Increment methods for the scope of score-based \textit{CURATE} algorithm. 
\newline
\textbf{Additive Increment Method:} This method increments the privacy budget for each iteration ($\epsilon_i$) as a function of current number of iterations ($i$), total assigned privacy budget ($\epsilon_{\text{Total}}$) and initial privacy budget ($\epsilon_0$). Mathematically, for every $i^{th}$ iteration, this method increments the privacy budget for each iteration as:
    $\epsilon_i = \epsilon_0(1+\frac{i}{I_{\text{add}}})$.
Given total privacy budget $\epsilon_{\text{Total}}$, initial privacy budget $\epsilon_0$, and number of iterations $I_{\text{add}}$, we can define $\epsilon_{\text{Total}}$ as:
\begin{align*}
    &\epsilon_{\text{Total}}=\frac{I_{\text{add}}}{2}\left[2\epsilon_0+(I_{\text{add}}-1)\frac{\epsilon_0}{I_{\text{add}}}\right]\\
    &I_{\text{add}}=\frac{\epsilon_{\text{Total}}+\frac{\epsilon_0}{2}}{\epsilon_0+\frac{\epsilon_0}{2}}.
\end{align*}
\textbf{Exponential Increment Method:} This method enables the increment of per iteration privacy budget as an exponential function of the initial budget ($\epsilon_0$) and current iteration $i$. For every $i^{th}$ iteration, the Exponential increment method defines the privacy budget as:
 $   \epsilon_i=\epsilon_0\cdot\exp\left(\frac{i}{I_\text{exp}}\right)$.
Given total privacy budget $\epsilon_{\text{Total}}$, initial privacy budget $\epsilon_0$, and possible number of iterations $I_{\text{exp}}$, we will define $\epsilon_{\text{Total}}$. 
\begin{align*}
    &\exp(0)\le \exp(1/I_{\text{exp}}) \le \ldots \le \exp(I_{\text{exp}}/I_{\text{exp}})\\
    &\sum_{i=0}^{I_{\text{exp}}}\exp\left(\frac{i}{I_{\text{exp}}}\right)\le I_{\text{exp}}\exp(1).
\end{align*}
To maintain the total privacy budget of $\epsilon_{\text{Total}}$, we can define the relationship between $I_{\text{exp}},\epsilon_{\text{Total}}$ and $\epsilon_0$ as:
\begin{align}
    &\epsilon_{\text{Total}}\ge I_{\text{exp}}\cdot\exp(1)\cdot\epsilon_0\nonumber\\
    &I_{\text{exp}} \le \frac{\epsilon_{\text{Total}}}{\epsilon_0 \cdot\exp(1)}.
\end{align}
\textbf{Multiplicative Increment Method:} This method enables the algorithm to increment the per iteration privacy budget ($\epsilon_i$) as a multiplicative function of the initial budget ($\epsilon_0$) and current iteration. For every $i^{th}$ iteration, the per-iteration privacy budget ($\epsilon_i$) is defined as: 
$   \epsilon_i=\epsilon_0^{(1+\frac{i}{I_{\text{mul}}})}$.
In this method, the possible number of iterations ($I_{\text{mul}}$) depends on the value of the factor $\epsilon_0^{1/I_{\text{mul}}}$. If, $\epsilon_0^{1/I_{\text{mul}}}\le 1$ then $\epsilon_0\le 1$ which indicates a high privacy regime else it indicates a low privacy regime where $\epsilon_0^{1/I_{\text{mul}}}\ge 1$ and $\epsilon_0\ge 1$. For the high privacy regime ($\epsilon_0\le 1$), we define total leakage $\epsilon_{\text{Total}}$ as:
\begin{align}
\epsilon_{\text{Total}}&=\frac{\epsilon_0(1-\epsilon_0^{\frac{1}{I_{\text{mul}}}\cdot I_{\text{mul}}})}{1-\epsilon_0^{1/I_{\text{mul}}}}=\frac{\epsilon_0(1-\epsilon_0)}{1-\epsilon_0^{1/I_{\text{mul}}}}\nonumber\\
    I_{\text{mul}}&=\frac{\log(\epsilon_0)}{\log\left(1-\frac{\epsilon_0(1-\epsilon_0)}{\epsilon_{\text{Total}}}\right)}.
\end{align}
For the case where the initial privacy budget $\epsilon_0>1$, we can derive the expression of $I_{\text{mul}}$ as:
\begin{align}
    \epsilon_{\text{Total}}&=\frac{\epsilon_0(\epsilon_0^{\frac{1}{I_{\text{mul}}}\cdot I_{\text{mul}}}-1)}{\epsilon_0^{\frac{1}{I_{\text{mul}}}}-1}=\frac{\epsilon_0(\epsilon_0-1)}{\epsilon_0^{\frac{1}{I_{\text{mul}}}}-1}\nonumber\\
    I_{\text{mul}}&=\frac{\log(\epsilon_0)}{\log\left(\frac{\epsilon_0(\epsilon_0-1)}{\epsilon_{\text{Total}}}+1\right)}.
\end{align}
This concludes the proof of Lemma \ref{sc_curate_lemma2}.

\bibliography{main}
\bibliographystyle{plain}
\end{document}